\documentclass[preprint,prd,aps,superscriptaddress]{revtex4-1}
\usepackage{hyperref}
\usepackage{bm}
\usepackage{amsmath}
\usepackage{amssymb}
\usepackage{graphicx}

\renewcommand{\Re}{\mathrm{Re}}
\renewcommand{\Im}{\mathrm{Im}}
\newcommand{\sign}{\mathrm{sign}}
\newcommand{\arctanh}{\mathrm{arctanh}}

\begin{document}

\title{One-loop pentagon integral with one offshell leg in $d$ dimensions from differential equations in $\epsilon$-form}

\author{Mikhail G.  Kozlov}
\affiliation{Budker Institute of  Nuclear Physics, Novosibirsk, 630090 Russia}
\affiliation{Novosibirsk State University, Novosibirsk, 630090 Russia}
\email{m.g.kozlov@inp.nsk.su}

\thanks{This work is supported by the RFBR grants No. 16-32-60033 and 16-02-00888}

\begin{abstract}
We apply differential equations technique to the calculation of the one-loop massless diagram with one offshell legs. Using reduction to $\epsilon$-form, we managed to obtain a simple one-fold integral representation exact in space-time dimensionality. Expansion of the obtained result in $\epsilon$ and analytical continuation to physical region are discussed.
\end{abstract}

\maketitle

\section{Introduction}

In this paper we consider the one-loop integral with massless internal lines and one off-shell leg  exactly in the dimension of space-time, which we call below \textit{the pentagon integral}. This type of integrals arises for example in a system of differential equations on one loop master integrals with massless internal lines and with more than five legs, for example on massless on-shell hexagon  integral. Also exact expression for the one-loop pentagon integral is needed to calculate the MHV multiloop amplitudes in the planar limit of perturbation theory, using Bern--Dixon--Smirnov (BDS) ansatz~\cite{BDS:2005}. 

The expression for pentagon integral has been known in the expansion in $ \epsilon $ up to $ \epsilon^0 $ for a long time~\cite{BDK:1994}. In Ref.~\cite{BDK:1994}, it was shown that up to $ \epsilon^0 $ order the pentagon integral in $ d=4-2\epsilon $ can be expressed via box integrals with one and two off shell legs. Also in~\cite{BDK:1994} it was shown that higher order terms are related to the expansion of the same pentagon integral in $ d=6-2\epsilon $ dimensions. 

In the previous paper~\cite{KL:2016} we considered  the pentagon integral with all on-shell legs exactly in $ d $ dimensions. In this paper we use the same algorithm of calculation as in~\cite{KL:2016}.

The result has a one-fold integral representation. We consider the pentagon integral in a subregion of Euclidean region and then perform the analytical continuation to all regions with real invariants.

In the second chapter, we introduce the notation and present the result for the Pentagon integral. The third chapter is dedicated to bringing the system of differential equations on the Pentagon to the e-form and obtain solutions of this system. In the fourth chapter, we discuss the analytic continuation to physical region of invariants. The fifth chapter is the conclusion. In Appendix A, we discuss the computation of the  triangle  master integral with all off-shell legs. In Appendix B and C we present the calculation of the easy box and the hard box master integrals respectively.

\section{Definitions and result}

The pentagon integral is defined as 
\begin{equation}
P^{\left(d\right)}\left(s_{1},\,s_{2},\,s_{3},\,s_{4},\,s_{5}\right)=\int\frac{d^{d}l}{i\,\pi^{d/2}\prod_{n=0}^{4}\left(l_{n}^{2}+i0\right)}\,,
\end{equation}
where 
\begin{equation}
l_{n}=l-\sum_{i=1}^{n}p_{i}\,,
\end{equation} and $p_i$ are the incoming momenta,
\begin{equation}
p_1^2=m_1^2=\mu_1\,,\quad p_{i}^{2}=0\;(i=2,\dots,5) \,,\quad\sum_{i=1}^{5}p_{i}=0\,,
\end{equation}
and the invariants $s_i$ are defined as
\begin{equation}
s_{n}=(p_{n-2}+p_{n+2})^2\,,\;\bm{s}\equiv(s_1,s_2,s_3,s_4,s_5,\mu_1).
\end{equation}
Here we adopt cyclic convention for indices, e.g. $s_{n\pm5}=s_{n}$. 
We find it convenient to use the following notation
\begin{equation}
\begin{split}
r_n&=-s_1\mu_1+\sum_{i=0}^4(-1)^is_{n+i}s_{n+i+1}\,,\quad n=1,\,3,\,4\,,\\
r_2&=-2\frac{\mu_1s_1s_2}{s_4}+s_1\mu_1+\sum_{i=0}^4(-1)^is_{2+i}s_{3+i}\,,\quad r_5=-2\frac{\mu_1s_1s_5}{s_3}+s_1\mu_1+\sum_{i=0}^4(-1)^is_{5+i}s_{6+i}\,,\\
\Delta&=\det\left(2p_{i}\cdot p_{j}|_{i,j=1,\ldots4}\right)\;,\quad S=\frac{4s_1s_2(s_3s_4-s_1\mu_1)s_5}{\Delta}\,,\\
\Delta_3&=4s_2s_5-(\mu_1-s_2-s_5)^2\;,\quad S_3=\frac{4s_2s_5\mu_1}{\Delta_3}\,.
\end{split}
\end{equation}

Using techniques described in detail in the next sections, we obtain the following result~\eqref{eq:result} exact in $ d $ representation for $ P^{(6-2\epsilon)} $ for real $ s_i $ in non-equivalent regions $ {\cal R}_i $. These regions defined by the list 
$\bigl(\sign(s_1),\sign(s_2),\sign(s_3),\sign(s_4),\sign(s_5),\sign(\mu_1)\bigr) $. Solution for other regions can be obtained using symmetry $ s_2\leftrightarrow s_5,\,s_3\leftrightarrow s_4 $ and the identity $ P^{(6-2\epsilon)}(\boldsymbol{s})=e^{i\pi\epsilon}\Bigl[P^{(6-2\epsilon)}(-\boldsymbol{s})\Bigr]^{*} $ following from Feynman parametrization.
\begin{equation}\label{eq:result}
\begin{split}
&P^{(6-2\epsilon)}(\boldsymbol{s})=\frac{2C(\epsilon)}{\epsilon}\biggl[-\sum_{i=1,3,4}(-s_i)^{-\epsilon}\Re\int_1^{\infty}\frac{dt\;t^{\epsilon-1}}{b_i(t)}\biggl(\arctan\frac{r_i}{b_i(t)}-\arctan\frac{g_i(t)}{b_i(t)}\biggr)+\\
&+(-s_2)^{-\epsilon}\Re\int_1^{\infty}\frac{dt\;t^{\epsilon-1}}{b_2(t)}\biggl(\arctan\frac{r_4}{b_2(t)}-\arctan\frac{g_2(t)}{b_2(t)}\biggr)+\\
&+(-s_5)^{-\epsilon}\Re\int_1^{\infty}\frac{dt\;t^{\epsilon-1}}{b_5(t)}\biggl(\arctan\frac{r_3}{b_5(t)}-\arctan\frac{g_5(t)}{b_5(t)}\biggr)-\\
&-(-\mu_1)^{-\epsilon}\Re\int_1^{\infty}\frac{dt\;t^{\epsilon-1}}{b_6(t)}\biggl(\arctan\frac{r_1}{b_6(t)}-\arctan\frac{g_6(t)}{b_6(t)}\biggr)+\\
&+\Theta(-s_2)\Theta(-s_5)\Theta(-\mu_1)\Bigl(\Theta(s_5-s_2-\mu_1)+\Theta(s_2-s_5-\mu_1)+\Theta(\mu_1-s_5-s_2)-2\Bigr)\times\\
&\times(-S_3)^{-\epsilon}\Re\int_1^{\infty}\frac{dt\;t^{\epsilon-1}}{b_0(t)}\biggl(\arctan\frac{g_-(t)}{b_0(t)}-\arctan\frac{g_+(t)}{b_0(t)}\biggr)+\frac{(-S)^{-\epsilon}}{\sqrt{\Delta}}\pi^{\frac{3}{2}}\frac{\Gamma(1/2-\epsilon)}{\Gamma(1-\epsilon)}\Theta(\Delta)B(\boldsymbol{s})\biggr]\,,
\end{split}
\end{equation}
\begin{table}
\small
\begin{tabular}{ccc}
 & region & $ B(\boldsymbol{s}) $\\
\hline
$ {\cal R}_1 $ & $ (------) $ & $ \Theta(s_3s_4-s_1\mu_1)\Bigl[\Theta(\mu_1-s_3)+\Theta(\mu_1-s_4)+\Theta(\sigma_5)+\Theta(\sigma_2)-2\Bigr] $\\
$ {\cal R}_{2} $ & $ (+-----) $ & $ 0 $\\
$ {\cal R}_{3} $ & $ (-+----) $ & $ \Theta(s_1\mu_1-s_3s_4)\Bigl[\Theta(\mu_1-s_3)-\Theta(\mu_1-s_4)+\Theta(S-S_3)\Theta(s_2s_3-s_3s_4+s_4s_5)-1\Bigr] $\\
$ {\cal R}_{4} $ & $ (--+---) $ & $ 0 $\\
$ {\cal R}_{5} $ & $ (-----+) $ & $ 1 $\\
$ {\cal R}_{6} $ & $ (++----) $ & $ \Theta(s_3-\mu_1)-\Theta(s_4-\mu_1)+\Theta(S-S_3)\Theta\bigl(s_3s_4(s_5-s_2)/\mu_1+s_2s_3-s_4s_5\bigr) $\\
$ {\cal R}_{7} $ & $ (+-+---) $ & $ \Theta(s_1\mu_1-s_3s_4)\Bigl[\Theta(s_4-\mu_1)-\Theta(\sigma_2)-\Theta(\sigma_5)+1\Bigr] $\\
$ {\cal R}_{8} $ & $ (+----+) $ & $ 0 $\\
$ {\cal R}_{9} $ & $ (-+---+) $ & $ 0 $\\
$ {\cal R}_{10} $ & $ (--+--+) $ & $ \Theta(s_3s_4-s_1\mu_1)+\Theta(s_1\mu_1-s_3s_4)\bigl(\Theta(s_3-\mu_1)-1\bigr) $\\
$ {\cal R}_{11} $ & $ (-++---) $ & $ -\Theta(s_1\mu_1-s_3s_4)-\Theta(s_3s_4-s_1\mu_1)\bigl(\Theta(s_4-\mu_1)-1\bigr)  $\\
$ {\cal R}_{12} $ & $ (--++--) $ & $ \Theta(s_3s_4-s_1\mu_1)\Bigl[\Theta(\sigma_2)+\Theta(\sigma_5)\Bigr] $\\
$ {\cal R}_{13} $ & $ (-+-+--) $ & $ -\Theta(s_3-\mu_1)+\Theta(S-S_3)\Theta\bigl(s_3s_4(s_5-s_2)/\mu_1+s_2s_3-s_4s_4\bigr) $\\
$ {\cal R}_{14} $ & $ (-+--+-) $ & $ \Theta(s_3s_4-s_1\mu_1)\Bigl[\Theta(s_3-\mu_1)+\Theta(s_4-\mu_1)\Bigr] $\\
$ {\cal R}_{15} $ & $ (+++---) $ & $ \Theta(s_3s_4-s_1\mu_1)\Theta(\mu_1-s_4)-\Theta(s_1\mu_1-s_3s_4) $\\
$ {\cal R}_{16} $ & $ (++-+--) $ & $ \Theta(s_3s_4-s_1\mu_1)\Bigl[-\Theta(\mu_1-s_3)+\Theta(S-S_3)\Theta(s_2s_3-s_3s_4+s_4s_5)\Bigr] $\\
$ {\cal R}_{17} $ & $ (++--+-) $ & $ 1 $\\
$ {\cal R}_{18} $ & $ (++---+) $ & $ \Theta(s_3s_4-s_1\mu_1)+\Theta(s_1\mu_1-s_3s_4)\Theta(s_2s_3-s_3s_4+s_4s_5) $\\
$ {\cal R}_{19} $ & $ (+-++--) $ & $ 0 $\\
$ {\cal R}_{20} $ & $ (+-+--+) $ & $ 0 $\\
\hline
\end{tabular} 
\caption{Coefficient $ B(\boldsymbol{s}) $ for non-equivalent regions $ {\cal R}_i $. Each region is marked by the list 
$\bigl(\sign(s_1),\sign(s_2),\sign(s_3),\sign(s_4),\sign(s_5),\sign(\mu_1)\bigr) $.}\label{tab:1}
\end{table}
where 
\begin{equation}
b_i(t)=\sqrt{\Delta(S\,t/s_i-1)}\;,\quad b_6(t)=\sqrt{\Delta(S\,t/\mu_1-1)}\,,\; b_0(t)=\sqrt{\Delta(S\,t/S_3-1)}\,,
\end{equation}
\begin{equation}
C(\epsilon)=\frac{\Gamma^2(1-\epsilon) \Gamma(1 + \epsilon)}{\Gamma(1-2\epsilon)}\,,
\end{equation}
\begin{equation}
\begin{split}
&g_1(t)=r_1+2s_2s_5(1-t)\,,\\
&g_3(t)=r_3+(1-t)\frac{2s_2(s_3s_4-s_1\mu_1)}{s_3-\mu_1}\,,\quad g_4(t)=r_4+(1-t)\frac{2s_5(s_3s_4-s_1\mu_1)}{s_4-\mu_1}\,,\\
&g_2(t)=r_4+\frac{(s_3s_4-s_1\mu_1)}{\mu_1}(s_2-s_5-\mu_1)\biggl[1-\sqrt{1+\frac{4s_5\mu_1}{(s_2-s_5-\mu_1)^2}(t-1)}\biggr]\,,
\end{split}
\end{equation}
\begin{equation}
\begin{split}
&g_5(t)=r_3+\frac{(s_3s_4-s_1\mu_1)}{\mu_1}(s_5-s_2-\mu_1)\biggl[1-\sqrt{1+\frac{4s_2\mu_1}{(s_5-s_2-\mu_1)^2}(t-1)}\biggr]\,,\\
&g_6(t)=r_1+s_1(\mu_1-s_2-s_5)\biggl[1-\sqrt{1+\frac{4s_2s_5}{(\mu_1-s_2-s_5)^2}(t-1)}\biggr]\,,\\
&g_{\pm}(t)=-s_2s_3+s_3s_4-s_4s_5\pm s_1\sqrt{\Delta_3(t-1)}\,,
\end{split}
\end{equation}
and factor $ B(\boldsymbol{s}) $  is shown in Table~\ref{tab:1}. By $ \Re\bigl(\arctan(\sqrt{x}/r)/\sqrt{x}\bigr) $ we understand the function
\begin{equation}
f(x,r)=
\begin{cases}
\frac{1}{\sqrt{x}}\arctan(\frac{\sqrt{x}}{r})\,, & x>0\\
\frac{1}{2\sqrt{-x}}\log\Bigl|\frac{r+\sqrt{-x}}{r-\sqrt{-x}}\Bigr|\,,& x<0
\end{cases}.
\end{equation}

In order to crosscheck our result, we have performed comparison with the results for the pentagon obtained using \texttt{Fiesta 4.1}~\cite{Fiesta4} (we compared only two orders in $ \epsilon $), and found perfect agreement. 

\section{Differential equations in $ \epsilon $-form}

In this section we consider integrals in $ d=4-2\epsilon $ dimensions in ``Euclidean'' region 
$$ s_1<0,\;s_2<0,\;s_3<0,\;s_4<0,\;s_5<0,\;\mu_1<0. $$

\begin{figure}
\includegraphics[width=\linewidth]{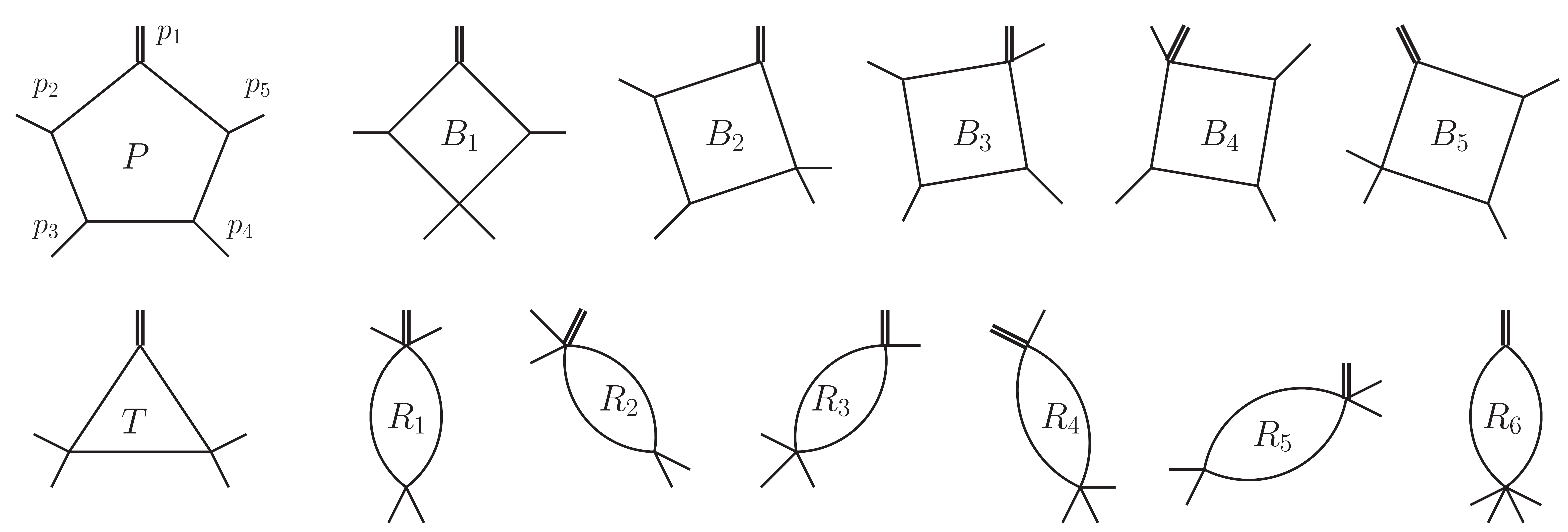}
\caption{Pentagon, $ B_1$ easy box, $ B_{2,5} $ hard boxes, $ B_{3,4} $ boxes, $ T $ triangle and bubble integrals.}\label{fig:1}
\end{figure}

We use integration by part (IBP) reduction, as implemented in \texttt{LiteRed} package, Ref.~\cite{LiteRed}, to obtain the system of partial differential equations for the pentagon integral $ P $ and twelve simpler master integrals, see Fig.~\ref{fig:1}.
Introducing the column-vector of master integrals
\begin{equation}
\bm{J}=\bigl(P,B_1,B_2,B_3,B_4,B_5,T,R_1,R_2,R_3,R_4,R_5,R_6\bigr)^T\,,
\end{equation}
we can represent the system of differential equations in the matrix form
\begin{equation}
\frac{\partial}{\partial s_i}\bm{J}=M_i(\bm{s},\epsilon)\bm{J}\;,\quad i=1,\dots,5\,;\quad
\frac{\partial}{\partial \mu_1}\bm{J}=M_6(\bm{s},\epsilon)\bm{J}\,,
\end{equation}
where $ M_i(\bm{s},\epsilon) $ are upper-triangular matrices of rational functions of $ s_j $, $ \mu_1 $ and $ \epsilon $.

We know the simpler master integrals, which are the bubbles
\begin{equation}
\begin{split}
&R_i=R(s_i)=\int\frac{d^dl}{i\pi^{d/2}l^2_{i+1}l^2_{i+3}}=\frac{C(\epsilon)}{\epsilon(1-2\epsilon)}(-s_i)^{-\epsilon}\,,\; i=1,\dots,5\,,\\
&R_6=R(\mu_1)=\int\frac{d^dl}{i\pi^{d/2}l^2l^2_{1}}=\frac{C(\epsilon)}{\epsilon(1-2\epsilon)}(-\mu_1)^{-\epsilon}\,,
\end{split}
\end{equation}
the triangle integral with all off-shell legs (see Appendix~\ref{sec:tri})
\begin{equation}
T=T(s_2,s_5,\mu_1)=\int\frac{d^dl}{i\pi^{d/2}l^2l^2_1l^2_3}\,,
\end{equation}
and the box integrals
\begin{equation}
B_i=\int\frac{d^dl}{i\pi^{d/2}\prod_{k=3}^6l^2_{i+k}}\,,
\end{equation}
where $ B_1 $ is the easy box (see appendix~\ref{sec:eb}), $ B_{2,5} $ are hard boxes (see appendix~\ref{sec:hb}) and $ B_{3,4} $ are boxes with one off-shell leg. The representation of the easy box and the boxes with one off-shell leg in terms of hypergeometric functions were obtained in~\cite{BDK:1994,DN:2001}. The representation of the box integral with one off-shell leg has the form
\begin{equation}\label{eq:box1m}
\begin{split}
&B(s,t,\mu)=\frac{2C(\epsilon)}{\epsilon^2s\,t}\biggl\{(-\mu)^{-\epsilon}\,_2F_1\Bigl(1,-\epsilon;1-\epsilon;\frac{\mu(s+t-\mu)}{s\,t}\Bigr)-\\
&-(-s)^{-\epsilon}\,_2F_1\Bigl(1,-\epsilon;1-\epsilon;\frac{s+t-\mu}{t}\Bigr)-(-t)^{-\epsilon}\,_2F_1\Bigl(1,-\epsilon;1-\epsilon;\frac{s+t-\mu}{s}\Bigr)\biggr\}\,.
\end{split}
\end{equation}
In order to reduce the system of differential equations to $ \epsilon $-form we find the appropriate basis
\begin{equation}\label{eq:basis}
\begin{split}
&P=\frac{C(\epsilon)}{\epsilon^2s_1s_2(s_3s_4-s_1\mu_1)s_5}\biggl(\sqrt{\Delta}\widetilde{P}+\frac{1}{2}\sum_{i=1}^{5}r_i\widetilde{B}_i\biggr)\,,\\
&B_1=\frac{C(\epsilon)}{\epsilon^2(s_3s_4-s_1\mu_1)}\widetilde{B}_1\,,\quad B_i=\frac{C(\epsilon)}{\epsilon^2s_{i+2}s_{i-2}}\widetilde{B}_i\,,i=1,\dots,5\,,\\
&T=\frac{C(\epsilon)}{\epsilon^2\sqrt{\Delta_3}}\widetilde{T}\,,\quad R_i=\frac{C(\epsilon)}{\epsilon(1-2\epsilon)}\widetilde{R}_i\,.
\end{split}
\end{equation}
The differential equations in the new basis can be written in $ d\log $ $ \epsilon $-form
\begin{equation}\label{eq:deq-p}
\begin{split}
d\widetilde{P}&=-\epsilon\biggl\{\widetilde{P}d\Bigl(\log S\Bigr)+\sum_{i=1}^5\widetilde{B}_id\Bigl(\arctanh a_i\Bigr)-\widetilde{T}d\Bigl(\arctan y\Bigr)+\\
&+2\sum_{i=1,3,4}\widetilde{R}_id\Bigl(\arctanh a_i-\arctanh a_{i+2}-\arctanh a_{i-2}\Bigr)+\\
&+\widetilde{R}_2d\Bigl(\arctanh a_2-2\arctanh a_4-\arctanh a_5\Bigr)+\\
&+\widetilde{R}_5d\Bigl(\arctanh a_5-2\arctanh a_3-\arctanh a_2\Bigr)+\\
&+\widetilde{R}_6d\Bigl(2\arctanh a_1+\arctanh a_2+\arctanh a_5\Bigr)\biggr\}\,;\\
d\widetilde{B}_1&=-\epsilon\biggl\{\widetilde{B}_1d\log(-S_{4}^{(1)})-2\widetilde{R}_1d\log\Bigl(\frac{S_{4}^{(1)}}{s_1}-1\Bigr)-2\widetilde{R}_6d\log\Bigl(\frac{S_{4}^{(1)}}{\mu_1}-1\Bigr)+\\
&+2\widetilde{R}_3d\log\Bigl(\frac{S_{4}^{(1)}}{s_3}-1\Bigr)+2\widetilde{R}_4d\log\Bigl(\frac{S_{4}^{(1)}}{s_4}-1\Bigr)\biggr\}\,;\\
d\widetilde{B}_2&=-\epsilon\Biggl\{\widetilde{B}_2d\log(-S_{4}^{(2)})-2\widetilde{T}d\arctan b_2+2\widetilde{R}_3d\log\Bigl(\frac{S_{4}^{(2)}}{s_3}-1\Bigr)-\\
&-\widetilde{R}_2d\log\frac{1+b_2^2}{(b_2-x_2)^2}+\widetilde{R}_6d\log\frac{1+b_2^2}{(b_2+x_6)^2}+\widetilde{R}_5d\log\frac{1+b_2^2}{(b_2+x_5)^2}\Biggr\}\,;\\
d\widetilde{B}_3&=-\epsilon\biggl\{\widetilde{B}_3d\log(-S_{4}^{(3)})-2\widetilde{R}_3d\log\Bigl(\frac{S_4^{(3)}}{s_3}-1\Bigr)+2\widetilde{R}_1d\log\Bigl(\frac{S_4^{(3)}}{s_1}-1\Bigr)+\\
&+2\widetilde{R}_5d\log\Bigl(\frac{S_4^{(3)}}{s_5}-1\Bigr)\biggr\}\,;\\
d\widetilde{T}&=-\epsilon\widetilde{T}d\log\Bigl(-S_3\Bigr)-2\epsilon \widetilde{R}_2d\arctan x_2-2\epsilon \widetilde{R}_5d\arctan x_5-2\epsilon \widetilde{R}_6d\arctan x_6\,;\\
d\widetilde{R}_i&=-\epsilon\widetilde{R}_id\bigl(\log s_i\bigr),\,i=1,\dots,5\,,\quad d\widetilde{R}_6=-\epsilon\widetilde{R}_6d\bigl(\log \mu_1\bigr)\,.
\end{split}
\end{equation}
We also need the equations for $ \widetilde{B}_4 $ and $\widetilde{B}_5  $, which have the same form as the  equations for $ \widetilde{B}_3 $ and $\widetilde{B}_2  $. In Eq.~\eqref{eq:deq-p} we use the following denotations
\begin{equation}
\begin{split}
&y=\frac{1}{\sqrt{\Delta\Delta_3}}\Bigl(-\mu_1^2s_1 + 2 \mu_1s_1s_2 - s_1s_2^2 - \mu_1s_2s_3 + s_2^2 s_3 + \mu_1s_3s_4 -s_2 s_3 s_4 + 2 \mu_1 s_1 s_5+\\
& + 2 \mu_1 s_2 s_5+ 2 s_1 s_2 s_5 - s_2 s_3 s_5 -\mu_1 s_4 s_5 - s_2 s_4 s_5 - s_3 s_4 s_5 - s_1 s_5^2 + s_4 s_5^2\Bigr)\,,\; a_i=\frac{r_i}{\sqrt{\Delta}}\,,\\
&S_4^{(1)}=\frac{s_3s_4-s_1\mu_1}{s_1+\mu_1-s_3-s_4}\,,\;S_4^{(2)}=-\frac{s_2s_3^2}{(s_3-\mu_1)(s_3-s_5)+s_2s_3}\,,\;S_4^{(3)}=\frac{s_1s_5}{s_3-s_1-s_5}\,,\\
&x_2=\frac{s_2-s_5-\mu_1}{\sqrt{\Delta}_3}\,,\quad x_5=\frac{s_5-s_2-\mu_1}{\sqrt{\Delta}_3}\,,\quad x_6=\frac{\mu_1-s_2-s_5}{\sqrt{\Delta}_3}\,,\\
&b_2=\frac{2s_2\mu_1}{s_4\sqrt{\Delta}_3}+x_5\,,\;b_5=\frac{2s_5\mu_1}{s_3\sqrt{\Delta}_3}+x_2\,.
\end{split}
\end{equation}
Let us fix the region of invariants $ {\cal R} $ where solutions for triangle  and hard boxes are more simple 
\begin{equation}\label{eq:region}
\begin{split}
{\cal R}=\{\boldsymbol{s}|\,&s_1<0,s_2<0,s_3<0,s_4<0,s_5<0,\mu_1<0,\\
&\Delta>0,s_3s_4-s_1\mu_1>0,\Delta_3>0,x_2>0,x_5<0,x_6>0\}\,.
\end{split}
\end{equation}
Here the signs of $ \Delta $ and $ s_3s_4-s_1\mu_1 $  fix the sign of $S$.

Let us now split the above differeintial system. Given a system $ d\widetilde{J} =dM\widetilde{J}$, we schematically depict the matrix $ dM $ by replacing each  nonzero element with ``$ * $''. For system~\eqref{eq:deq-p} we have:
\begin{equation}
dM=\left[\tiny\begin{array}{*{13}c}
*&*&*&*&*&*&*&*&*&*&*&*&*\\
0&*&0&0&0&0&0&*&0&*&*&0&*\\
0&0&*&0&0&0&*&0&*&*&0&*&*\\
0&0&0&*&0&0&0&*&0&*&0&*&0\\
0&0&0&0&*&0&0&*&*&0&*&0&0\\
0&0&0&0&0&*&*&0&*&0&*&*&*\\
0&0&0&0&0&0&*&0&*&0&0&*&*\\
0&0&0&0&0&0&0&*&0&0&0&0&0\\
0&0&0&0&0&0&0&0&*&0&0&0&0\\
0&0&0&0&0&0&0&0&0&*&0&0&0\\
0&0&0&0&0&0&0&0&0&0&*&0&0\\
0&0&0&0&0&0&0&0&0&0&0&*&0\\
0&0&0&0&0&0&0&0&0&0&0&0&*
\end{array}
\right] \,.
\end{equation}
Then we interpret this schematic form as adjacency matrix of the directed graph, with ``$*_{ij}$'' denoting a directed edge $ i \to j$. In general, the node $i$ is said to be an \textit{ancestor} of the node $j$ if there is a directed path from $i$ to $j$. A \textit{leaf} is a node which is not an ancestor of any other node. To each leaf we associate the subgraph consisting of the leaf itself and of all its ancestors.
For each such subgraph, we search for a solution of the original system having the form of the column vector with zeros put in all entries except the ones corresponding to the nodes of the subgraph~\cite{KL:2016}.
The general solution of differential system is written as the sum over different leaves. Here we have six leaves, $ R_i;\,i=1,\dots,6 $.  We search for the solution in the form
\begin{equation}\label{eq:sumJ}
\widetilde{\bf J}=\sum_{i=1}^6\widetilde{\bf J}^{(i)}\,,
\end{equation}
where
\begin{equation}
\begin{split}
&\widetilde{\bf J}^{(1)}=\bigl(\widetilde{P}^{(1)},\widetilde{B}_1^{(1)},0,\widetilde{B}_3^{(1)},\widetilde{B}_4^{(1)},0,0,\widetilde{R}_1,0,0,0,0,0\bigr)^T\,,\\
&\widetilde{\bf J}^{(2)}=\bigl(\widetilde{P}^{(2)},0,\widetilde{B}_2^{(2)},0,\widetilde{B}_4^{(2)},\widetilde{B}_5^{(2)},\widetilde{T}^{(2)},0,\widetilde{R}_2,0,0,0,0\bigr)^T\,,\\
&\widetilde{\bf J}^{(3)}=\bigl(\widetilde{P}^{(3)},\widetilde{B}_1^{(3)},0,\widetilde{B}_3^{(3)},0,\widetilde{B}_5^{(3)},0,0,0,\widetilde{R}_3,0,0,0\bigr)^T\,,\\
&\widetilde{\bf J}^{(4)}=\bigl(\widetilde{P}^{(4)},\widetilde{B}_1^{(4)},\widetilde{B}_2^{(4)},0,\widetilde{B}_4^{(4)},0,0,0,0,0,\widetilde{R}_4,0,0\bigr)^T\,,\\
&\widetilde{\bf J}^{(5)}=\bigl(\widetilde{P}^{(5)},0,\widetilde{B}_2^{(5)},\widetilde{B}_3^{(5)},0,\widetilde{B}_5^{(5)},\widetilde{T}^{(5)},0,0,0,0,\widetilde{R}_5,0\bigr)^T\,,\\
&\widetilde{\bf J}^{(6)}=\bigl(\widetilde{P}^{(6)},\widetilde{B}_1^{(6)},\widetilde{B}_2^{(6)},0,0,\widetilde{B}_5^{(6)},\widetilde{T}^{(6)},0,0,0,0,0,\widetilde{R}_6\bigr)^T\,.
\end{split}
\end{equation}

\subsection*{Equation for $ \widetilde{P}^{(k)} $, $ k=1,3,4 $.}
Let us consider the differential equation for $ \widetilde{P}^{(k)} $ with $ k=1,3,4 $. This equations is the same as for the massless pentagon case~\cite{KL:2016} (in that work we had a slightly different basis):
\begin{equation}
\begin{split}
d\widetilde{P}^{(k)}=-\epsilon\biggl\{\widetilde{P}^{(k)}d\Bigl(\log S\Bigr)&+\sum_{i=k,k\pm2}\widetilde{B}^{(k)}_id\Bigl(\arctanh a_i\Bigr)+\\
&+2\widetilde{R}_kd\Bigl(\arctanh a_k-\arctanh a_{k+2}-\arctanh a_{k-2}\Bigr)\biggr\}\,.
\end{split}
\end{equation}
From Eq.~\eqref{eq:box1m} we can  identify $ \widetilde{B}_i^{(k)} $:
\begin{equation}
\begin{split}
&\widetilde{B}_i^{(k)}=2(-1)^{(k-i)/2}(-s_k)^{-\epsilon}\Re\,_2F_1\biggl(1,-\epsilon;1-\epsilon;\frac{s_k}{S}\Bigl(1-a_i^2\Bigr)\biggr)=\\
&=2(-1)^{(k-i)/2}(-s_k)^{-\epsilon}\Biggl\{1-\epsilon(1-a_i^2)\Re\int_1^{\infty}\frac{dt\,t^{\epsilon-1}}{t\,S/s_k-1+a_i^2+i0}\Biggr\}\,.
\end{split}
\end{equation}
Using the integral representation for $ \widetilde{B}_i^{(k)} $ we arrive at the following differential equation for $ \widetilde{P}^{(k)} $:
\begin{equation}\label{eq:1}
\begin{split}
&d\Bigl((-S)^{\epsilon}\widetilde{P}^{(k)}\Bigr)=H_k^{(k)}da_k+H_{k+2}^{(k)}da_{k+2}+H_{k-2}^{(k)}da_{k-2}\,,\\
&H_i^{(k)}=2\epsilon^2(-1)^{(k-i)/2}\Bigl(\frac{S}{s_k}\Bigr)^{\epsilon}\int_1^{\infty}\frac{dt\,t^{\epsilon-1}}{t\,S/s_k-1+a_i^2}\,,\\
&\frac{S}{s_k}=1+a_{k-2}a_{k+2}-a_k(a_{k-2}+a_{k+2})\,,\;k=1,3,4.
\end{split}
\end{equation}
The right-hand side of~\eqref{eq:1} depends only on free dimensionless variables $ a_n,\,(n=i,i\pm2) $. It is easy to check that~\eqref{eq:1} is a total differential. Then from the differential equation we have (as in the article~\cite{KL:2016})
\begin{equation}
(-S)^{\epsilon}\widetilde{P}^{(k)}=\int_{-\sigma_k\infty}^{a_k}H_k^{(k)}(a,a_{k+2},a_{k-2})da+g(a_{k+2},a_{k-2},\epsilon)\,,
\end{equation}
where $ \sigma_k=\sign(a_{k+2}+a_{k-2}) $. Next,  we can check that $ g $ depends only on $ \epsilon $:
\begin{equation}
\frac{\partial g(a_{k+2},a_{k-2},\epsilon)}{\partial a_{k\pm2}}=H^{(k)}_{k\pm2}(a_k\rightarrow -\sigma_k\infty)=\frac{2\epsilon^2}{\epsilon-1}\bigl(-a_k(a_{k+2}+a_{k-2})\bigr)^{\epsilon-1}\Big|_{a_k\rightarrow-\sigma_k\infty}=0\,.
\end{equation}
Substituting the explicit form of $ H_i^{(i)} $, we have
\begin{equation}
\begin{split}
&(-S)^{\epsilon}\widetilde{P}^{(k)}=2\epsilon^2\Re\int_{-\sigma_k\infty}^{a_k}da\int_1^{\infty}\bigl(K(a)\bigr)^{\epsilon}\frac{dt\,t^{\epsilon-1}}{K(a) t-1+a^2+i0}+g(\epsilon)\,,\\
&K(a)=1+a_{k-2}a_{k+2}-a(a_{k-2}+a_{k+2})\,.
\end{split}
\end{equation}
Note that $ K(a)>0 $ in the whole integration domain. After making the substitution $ t\rightarrow t/K(a) $, changing integration order  we have and integrating over $ a $ we have
\begin{equation}\label{eq:P1}
\widetilde{P}^{(k)}=2\epsilon^2(-s_k)^{-\epsilon}\sqrt{\Delta}\Re\int_1^{\infty}\frac{dt\,t^{\epsilon-1}}{b_k(t)}\biggl\{\arctan\frac{r_k}{b_k(t)}-\arctan\frac{g_k(t)}{b_k(t)}\biggr\}\,,
\end{equation}
where 
\begin{equation}\label{eq:b}
b_k(t)=\sqrt{\Delta\Bigl(\frac{S}{s_k}t-1\Bigr)}\,,\quad g_k(t)=r_k+\frac{S\Delta}{s_k(r_{k+2}+r_{k-2})}(1-t)\,.
\end{equation}

\subsection*{Equation for $ \widetilde{P}^{(j)} $, $ j=2,\,5,\,6 $.}
Recall that we are working in the region~\eqref{eq:region}. The equation for $ \widetilde{P}^{(2)} $ (and equations for  $ \widetilde{P}^{(5)} $, $ \widetilde{P}^{(6)} $ ) is more complicated than the  equation for $ \widetilde{P}^{(1)}$ and has four independent variables $ a_2,\,a_5,\,a_4 $ and $ y $:
\begin{equation}\label{eq:deq2}
\begin{split}
&d\widetilde{P}^{(2)}=-\epsilon\biggl\{\widetilde{P}^{(2)}d\Bigl(\log S\Bigr)+\widetilde{B}_2^{(2)}d\Bigl(\arctanh a_2\Bigr)+\widetilde{B}_5^{(2)}d\Bigl(\arctanh a_5\Bigr)+\widetilde{B}_4^{(2)}d\Bigl(\arctanh a_4\Bigr)-\\
&-\widetilde{T}^{(2)}d\Bigl(\arctan y\Bigr)+\widetilde{R}_2d\Bigl(\arctanh a_2-2\arctanh a_4-\arctanh a_5\Bigr)\biggr\}\,,
\end{split}
\end{equation}
where from equations~\eqref{eq:box1m},~\eqref{eq:HBc} and~\eqref{eq:T} we have
\begin{equation}\label{eq:2}
\begin{split}
&\widetilde{T}^{(2)}=-2\epsilon(-S_3)^{-\epsilon}\Re\int_{+\infty}^{x_2}(1+z^2)^{\epsilon-1}dz\,,\quad S_3=s_2(1+x_2^2)\,,\\
&\widetilde{B}_2^{(2)}=2\epsilon(-S_3)^{-\epsilon}\Re\int_{+\infty}^{x_2}\frac{(1+z^2)^{\epsilon}}{b_2+z+i0}\,,\;\widetilde{B}_5^{(2)}=2\epsilon(-S_3)^{-\epsilon}\Re\int_{+\infty}^{x_2}\frac{(1+z^2)^{\epsilon}}{b_5-z+i0}\,,\\
&\widetilde{B}^{(2)}_4=2(-s_2)^{-\epsilon}\Re\,_2F_1\Bigl(1,-\epsilon;1-\epsilon;\frac{s_2}{S}(1-a_4^2)\Bigr)=\\
&=2(-s_2)^{-\epsilon}\Biggl\{1-\epsilon\Re\int_1^{\infty}\frac{(1-a_4^2)\,dt\,t^{\epsilon-1}}{t\,S/s_2-1+a_4^2+i0}\Biggr\}\,.
\end{split}
\end{equation}
The native variables for the hard boxes and the triangle are $ b_2,\,b_5 $ and $ x_2 $. We have to express them through independent variables $ a_2,\,a_4,\,a_5 $ and $ y $. From equalities
\begin{equation}\label{eq:4}
\frac{S}{s_2}=(1+x_2^2)\frac{1-a^2_2}{1+b_2^2}=(1+x_2^2)\frac{1-a^2_5}{1+b_5^2}\,,\quad y=\frac{b_2b_5-a_2a_5}{b_2a_5+a_2b_5}\;,\quad\frac{b_2+x_2}{b_5-x_2}=\frac{a_2-a_4}{a_5+a_4}
\end{equation}
we can express $ b_2,\,b_5,\,x_2 $:
\begin{equation}
\begin{split}
&b_2=a_2y+a_5\sqrt{(1+y^2)\frac{1-a_2^2}{1-a_5^2}}\,,\quad b_5=a_5y+a_2\sqrt{(1+y^2)\frac{1-a_5^2}{1-a_2^2}}\,,\\
&x_2=-a_4y+(a_2-a_4-a_5+a_2a_4a_5)\sqrt{\frac{1+y^2}{(1-a_2^2)(1-a_5^2)}}\,.
\end{split}
\end{equation}
The relationship between variables $ b_2,\,b_5,\,x_2 $ and $ a_2,\,a_4,\,a_5,\,y $ can be obtained using Gr\"obner basis. Another way to derive the relationship is to use relations~\eqref{eq:4}. 

Equations for $ \widetilde{P}^{(5)} $ and $ \widetilde{P}^{(6)} $ can be found from the equation for $ \widetilde{P}^{(2)} $. The equation for $ \widetilde{P}^{(5)} $ we obtain by replacing $ a_2\leftrightarrow a_5 $, $ a_4\rightarrow a_3 $ (or $ x_2\rightarrow x_5 $, $ b_2\leftrightarrow b_5 $) in~\eqref{eq:deq2}. The equation for $ \widetilde{P}^{(6)} $ we derive by replacing $ \widetilde{P}^{(2)}\rightarrow-\widetilde{P}^{(6)} $, $ y\rightarrow -y $, $ a_2\rightarrow -a_2 $ and $ a_4\rightarrow a_1 $ (or $ x_2\rightarrow x_6 $, $ b_5\rightarrow -b_5 $) in~\eqref{eq:deq2}. Using expressions~\eqref{eq:2} we arrive at the following differential equation for $ \widetilde{P}^{(2)} $:
\begin{equation}
\begin{split}
&d\Bigl((-S)^{\epsilon}\widetilde{P}^{(2)}\Bigr)=H_1dy+H_2da_2+H_4da_4+H_5da_5\,,\\
&H_1=-\frac{2\epsilon^2}{1+y^2}\biggl(\frac{1-a_2^2}{1+b_2^2}\biggr)^{\epsilon}\Re\int_{+\infty}^{x_2}(1+z^2)^{\epsilon-1}dz\,,\\
&H_2=-2\epsilon^2\frac{(1-a_2^2)^{\epsilon-1}}{(1+b_2^2)^{\epsilon}}\Biggl\{\frac{(1+x_2^2)^{\epsilon}}{2\epsilon}+\Re\int_{+\infty}^{x_2}\frac{(1+z^2)^{\epsilon}dz}{b_2+z+i0}\Biggr\}\,,\\
&H_4=-2\epsilon^2\biggl((1+x_2^2)\frac{1-a_2^2}{1+b_2^2}\biggr)^{\epsilon-1}\Re\int_1^{\infty}\frac{t^{\epsilon-1}dt}{t-\frac{(1-a_4^2)(1+b_2^2)}{(1-a_2^2)(1+x_2^2)}+i0}\,,\\
&H_5=-2\epsilon^2\frac{(1-a_5^2)^{\epsilon-1}}{(1+b_5^2)^{\epsilon}}\Biggl\{\frac{(1+x_2^2)^{\epsilon}}{2\epsilon}+\Re\int_{+\infty}^{x_2}\frac{(1+z^2)^{\epsilon}dz}{b_5-z+i0}\Biggr\}\,.
\end{split}
\end{equation}
Let us integrate equation $ \frac{\partial }{\partial a_4}\bigl((-S)^{\epsilon}\widetilde{P}_2\bigr)=H_4 $ (as simplest equation):
\begin{equation}
(-S)^{\epsilon}\widetilde{P}^{(2)}=\int_{+\infty}^{a_4}H_4(a_2,z,a_5,y)dz+g(a_2,a_5,y,\epsilon)
\end{equation}
where $ g(a_2,a_5,y,\epsilon) $ is some function to be fixed. It is easy to check that $ g $ depends only on $ \epsilon$:
\begin{equation}
\frac{\partial g(a_2,a_5,y,\epsilon)}{\partial a_{2,5}}=H_{2,5}(a_4\rightarrow+\infty)=0\,,\quad\frac{\partial g(a_2,a_5,y,\epsilon)}{\partial y}=H_{1}(a_4\rightarrow+\infty)=0\,.
\end{equation}
The limit $ a_4\rightarrow +\infty $ corresponds to the limit $ x_2\rightarrow+\infty $ 
\begin{equation}
x_2\Big|_{a_4\rightarrow+\infty}\approx\frac{(-\mu_1)\sqrt{\Delta}}{(s_3s_4-\mu_1s_1)\sqrt{\Delta_3}}a_4\rightarrow+\infty\,.
\end{equation}
Substituting the explicit form of $ H_4 $,  we have
\begin{equation}
(-S)^{\epsilon}\widetilde{P}^{(2)}=-2\epsilon^2\int_{+\infty}^{a_4}dz\Re\int_1^{\infty}\frac{\bigl(t\,K(z)\bigr)^{\epsilon-1}dt}{t-(1-z^2)/K(z)+i0}+g(\epsilon)\,.
\end{equation}
Here we introduce function $ K(z) $ (this function is an analog of function $ K(a) $ in the~\cite{KL:2016}):
\begin{equation}
\begin{split}
&K(z)=K(z_0)\bigl(1+A^2(z-z_0)^2\bigr)\,,\quad K(a_4)\equiv\frac{S}{s_2}=\frac{1-a_2^2}{1+b_2^2}(1+x_2^2)\,,\\
&K(z_0)=\frac{1-a_2^2}{1+b_2^2}=\frac{s_1(s_3s_4-\mu_1s_1)\Delta_3}{\mu_1\Delta}\,,\quad A=\frac{b_2+b_5}{a_2+a_5}=\frac{\mu_1\sqrt{\Delta}}{(s_3s_4-\mu_1s_1)\sqrt{\Delta_3}}\,,\\
&x_2=-A(a_4-z_0)\,,\quad z_0=\frac{a_2b_5-a_5b_2}{b_2+b_5}=\frac{s_3 s_4 (s_2-s_5) + \mu_1(s_4s_5 - s_2s_3)}{\mu_1\sqrt{\Delta}}\,.
\end{split}
\end{equation}
Note that $ K(z)>0 $ in the whole integration domain. Making substitution $ t\rightarrow t/K(z) $ and changing the order of integration we have
\begin{equation}\label{eq:3}
\begin{split}
&\widetilde{P}^{(2)}=2\epsilon^2(-S)^{-\epsilon}\int_{K(a_4)}^{\infty}dt\,t^{\epsilon-1}\Re\int_{K_+^{-1}(t)}^{a_4}\frac{dz}{t-1+z^2+i0}+g(\epsilon)(-S)^{-\epsilon}\,,\\
&K_{\pm}^{-1}(t)=z_0\pm\sqrt{\frac{1}{A^2}\biggl[\frac{t}{K(z_0)}-1\biggr]}\,.
\end{split}
\end{equation}
After integration~\eqref{eq:3} over $ z $ and change of variable $ t\rightarrow t K(a_4) $  we obtain
\begin{equation}\label{eq:P2}
\begin{split}
\widetilde{P}^{(2)}&=2\epsilon^2(-s_2)^{-\epsilon}\Re\int_1^{\infty}\frac{dt\,t^{\epsilon-1}}{\sqrt{K(a_4)t-1}}\biggl(\arctan\frac{a_4}{\sqrt{K(a_4)t-1}}-\\
&-\arctan\frac{a_4+x_2\bigl(1-\sqrt{t+(t-1)/x_2^2}\bigr)/A}{\sqrt{K(a_4)t-1}}\biggr)+g(\epsilon)(-S)^{-\epsilon}=\\
&=2\epsilon^2(-s_2)^{-\epsilon}\sqrt{\Delta}\Re\int_1^{\infty}\frac{dt\,t^{\epsilon-1}}{b_2(t)}\biggl(\arctan\frac{r_4}{b_2(t)}-\arctan\frac{g_2(t)}{b_2(t)}\biggr)+g(\epsilon)(-S)^{-\epsilon}\,,\\
&b_2(t)=\sqrt{\Delta\Bigl(\frac{S}{s_2}t-1\Bigr)}\,,\quad g_2(t)=r_4+x_2\frac{r_2+r_5}{b_2+b_5}\biggl(1-\sqrt{t+\frac{t-1}{x_2^2}}\biggr)\,,
\end{split}
\end{equation}
where $ b_2(t) $ is defined in~\eqref{eq:b}. It is easy to see that this solution has the same form as $ \widetilde{P}^{(1)} $ (
the  difference is only in the function $ g_2(t) $, which is not linear with respect to $ t $). Let us represent the solution in the form
\begin{equation}
\widetilde{P}^{(2)}=(-s_2)^{-\epsilon}F(a_2,a_5,a_4,y)\,.
\end{equation}
Then solutions for $ \widetilde{P}^{(5)} $ and $ \widetilde{P}^{(6)} $ are
\begin{equation}
\widetilde{P}^{(5)}=(-s_5)^{-\epsilon}F(a_5,a_2,a_3,y)\,,\quad\widetilde{P}^{(6)}=-(-\mu_1)^{-\epsilon}F(-a_2,a_5,a_1,-y)\,.
\end{equation}
Solution for $ \widetilde{P}^{(5)} $ we can be obtained by replacing $ s_2\leftrightarrow s_5 $ and $ s_3\leftrightarrow s_4 $ in~\eqref{eq:P2}. For $ \widetilde{P}^{(6)} $ replacement $ a_4\rightarrow a_1 $, $ a_2\rightarrow-a_2 $ and $ y\rightarrow -y $ corresponds to $ x_2\rightarrow x_6 $ and $ b_5\rightarrow -b_5 $.

\subsection*{Constant $ g(\epsilon) $}
Using equations~\eqref{eq:basis},~\eqref{eq:sumJ} we can write the solution for the pentagon integral in the form
\begin{equation}\label{eq:pd4}
P=\frac{C(\epsilon)\sqrt{\Delta}}{\epsilon^2s_1s_2(s_3s_4-s_1\mu_1)s_5}\biggl(\sum_{i=1}^6\widetilde{P}^{(i)}+g(\epsilon)(-S)^{-\epsilon}\biggr)+\frac{r_1}{2s_5s_1s_2}B_1+\frac{s_3s_4}{s_3s_4-s_1\mu_1}\sum_{i=2}^5\frac{r_i}{2s_{i-1}s_is_{i+1}}B_i\,.
\end{equation}
Let us fix the constant $ g(\epsilon) $. Note that the condition $ \Delta=0 $ implies the existence of a linear relation between $ p_1,\dots,p_4 $, therefore  we can express the pentagon integral at $ \Delta=0 $ in terms of the box integrals. Moreover, $ \Delta=0 $ is not a branching ponit of $ P $. The only way to satisfy these two conditions is to require that 
\begin{equation}\label{eq:5}
\sum_{i=1}^6\widetilde{P}^{(i)}+g(\epsilon)(-S)^{-\epsilon}\Big|_{\Delta\rightarrow 0}\rightarrow 0\,.
\end{equation}
Consider limit $ \Delta\rightarrow 0 $ in the case $ s_1\rightarrow 0 $ with condition $ s_2s_3-s_3s_4+s_4s_5=0 $. In the limit $ s_1\rightarrow 0 $ we have
\begin{equation}
\begin{split}
&\widetilde{P}^{(1)}\sim\widetilde{P}^{(5)}\sim\widetilde{P}^{(6)}\sim\Delta^{\frac{1}{2}-\epsilon}\rightarrow 0\,,\;\widetilde{P}^{(2)}\rightarrow 2\epsilon^2\pi^{\frac{3}{2}}\frac{\Gamma(1/2-\epsilon)}{\Gamma(1-\epsilon)}\,,\\
&\widetilde{P}^{(3)}\rightarrow -2\epsilon^2\pi^{\frac{3}{2}}\frac{\Gamma(1/2-\epsilon)}{\Gamma(1-\epsilon)}\Theta(\mu_1-s_3)\,,\quad\widetilde{P}^{(4)}\rightarrow -2\epsilon^2\pi^{\frac{3}{2}}\frac{\Gamma(1/2-\epsilon)}{\Gamma(1-\epsilon)}\Theta(\mu_1-s_4)\,.
\end{split}
\end{equation}
Therefore from Eq.~\eqref{eq:5} we obtain
\begin{equation}
g(\epsilon)=2\epsilon^2\pi^{\frac{3}{2}}\frac{\Gamma(1/2-\epsilon)}{\Gamma(1-\epsilon)}\Bigl(\Theta(\mu_1-s_3)+\Theta(\mu_1-s_4)-1\Bigr)\,.
\end{equation}
\subsection*{Solution for $ d=6-2\epsilon $}
Let us consider now the dimensional recurrence relation
\begin{equation}\label{eq:drr}
P^{(6-2\epsilon)}=\frac{s_1s_2(s_3s_4-s_1\mu_1)s_5}{\epsilon\Delta}\Biggl(P^{(4-2\epsilon)}-\frac{r_1}{2s_1s_2s_5}B_1^{(4-2\epsilon)}-\frac{s_3s_4}{s_3s_4-s_1\mu_1}\sum_{i=2}^5\frac{r_i}{2s_{i-1}s_{i}s_{i+1}}B_i^{(4-2\epsilon)}\Biggr)\,.
\end{equation}
This relation can be easily obtained with the \texttt{LiteRed}~\cite{LiteRed}. Comparing~\eqref{eq:pd4} and~\eqref{eq:drr}, we get
\begin{equation}\label{eq:pentagonD6}
P^{(6-2\epsilon)}=\frac{C(\epsilon)}{\epsilon^3\sqrt{\Delta}}\biggl[\sum_{i=1}^6\widetilde{P}^{(i)}+2\epsilon^2\pi^{\frac{3}{2}}\frac{\Gamma(1/2-\epsilon)}{\Gamma(1-\epsilon)}\Bigl(\Theta(\mu_1-s_3)+\Theta(\mu_1-s_4)-1\Bigr)\biggr]\,.
\end{equation}
Next, we write the solution for the pentagon in $ d=6-2\epsilon  $ in the region $ {\cal R} $~\eqref{eq:region}
\begin{equation}
P^{(6-2\epsilon)}=\frac{2C(\epsilon)}{\epsilon}\biggl[\sum_{i=1}^6\widehat{P}_i-{\cal H}(\boldsymbol{s},\epsilon)\Bigl(\Theta(\mu_1-s_3)+\Theta(\mu_1-s_4)-1\Bigr)\biggr]\,,
\end{equation}
where $ {\cal H}(\boldsymbol{s},\epsilon) $ is the homogenius solution of the differential equation for the pentagon:
\begin{equation}\label{eq:Hsol}
{\cal H}(\boldsymbol{s},\epsilon)=\pi^{\frac{3}{2}}\frac{\Gamma(1/2-\epsilon)}{\Gamma(1-\epsilon)}\frac{(-S)^{-\epsilon}}{\sqrt{\Delta}}\,,
\end{equation}
and $ \widehat{P}_i $ are
\begin{equation}\label{eq:hatp}
\begin{split}
&\widehat{P}_i=(-1)^{\alpha_i}(-s_i)^{-\epsilon}\Re\int_1^{\infty}\frac{dt\,t^{\epsilon-1}}{b_i(t)}\biggl[\arctan\frac{r_{k(i)}}{b_i(t)}-\arctan\frac{g_i(t)}{b_i(t)}\biggr]\,,\\
&\alpha_{i}=
\begin{cases}
0 & i=2,\,5\\
1 & i=1,\,3,\,4,\,6
\end{cases}
\,,\quad k(i)=i,\,i=1,3,4;\;k(2)=4,\;k(5)=3,\,k(6)=1\,,\\
\end{split}
\end{equation}
\begin{equation}\label{eq:gi}
\begin{split}
&g_i(t)=r_i+\frac{4s_1s_2(s_3s_4-s_1\mu_1)s_5}{s_i(r_{i+2}+r_{i-2})}(1-t)\,,\quad i=1,3,4\;,\\
&g_2(t)=r_4+\frac{(s_3s_4-s_1\mu_1)(s_2-s_5-\mu_1)}{\mu_1}\biggl[1-\sqrt{1+\frac{4s_5\mu_1}{(s_2-s_5-\mu_1)^2}(t-1)}\biggr]\,,\\
&g_5(t)=r_3+\frac{(s_3s_4-s_1\mu_1)(s_5-s_2-\mu_1)}{\mu_1}\biggl[1-\sqrt{1+\frac{4s_2\mu_1}{(s_5-s_2-\mu_1)^2}(t-1)}\biggr]\,,\\
&g_6(t)=r_1+s_1(\mu_1-s_2-s_5)\biggl[1-\sqrt{1+\frac{4s_2s_5}{(\mu_1-s_2-s_5)^2}(t-1)}\biggr]\,.
\end{split}
\end{equation}
Here we assume that $ s_6\equiv\mu_1 $ and $ b_i(t)=\sqrt{\Delta\bigl(S\,t/s_i-1\bigr)} $.

\section{Analytical continuation}

The real part of the integrals~\eqref{eq:hatp} is equivalent to half of the sum of the integrals over a contour passing above the singular points and the contour passing under the singular points. Let us introduce a new representation for integrand~\eqref{eq:hatp}:
\begin{equation}\label{eq:newrep}
\begin{split}
&\widehat{P}_i=(-s_i)^{-\epsilon}\Re\int_1^{\infty}dt\,t^{\epsilon-1}W_i(\boldsymbol{s},t)\\
&\Re W_i(\boldsymbol{s},t)=\frac{1}{2}\sum_{\pm}\biggl\{\frac{1}{r_{k(i)}}f\biggl(\frac{r_{k(i)}^2}{\Delta\frac{S}{s_i}(t\pm i0)-\Delta}\biggr)-\frac{1}{g_i(t\pm i0)}f\biggl(\frac{g_i(t\pm i0)^2}{\Delta\frac{S}{s_i}(t\pm i0)-\Delta}\biggr)\biggr\}
\end{split}
\end{equation}
where  function $ f(z)=\sqrt{z}\arctan(\sqrt{z}) $ is defined on the complex plane with a cut going from $ -\infty  $ to $ -1 $. The Riemann surface, corresponding to the multivaluated function $ F(z) $~\cite{KL:2016} with the main branch defined by $ F^{(0)}(z)=f(z) $, is glued of set of sheets numbered by $ n\in\mathbb{Z} $ with two cuts. On the $ n $-th sheet the function is defined as
\begin{equation}
\begin{split}
&F^{(n)}(z)=-\frac{\sqrt{-z}}{2}\ln\frac{1+\sqrt{-z}}{1-\sqrt{-z}}+i\pi n\sqrt{-z}\,,\;n\in\mathbb{Z}\,,\\
&F^{(n)}(x\pm i0)=
\begin{cases}
\sqrt{x}\arctan\sqrt{x}\pm\pi n\sqrt{z}=F^{(-n)}(x\mp i0), &x>0\\
-\frac{1}{2}\sqrt{-x}\ln\frac{1+\sqrt{-x}}{\sqrt{-x}-1}+i\pi(n\pm1/2)\sqrt{-x}=F^{(n\pm1)}(x\mp i0), & x<-1
\end{cases}\,.
\end{split}
\end{equation}
The integrand of~\eqref{eq:newrep} has the following branching points on the real axis of $ t $:
\begin{itemize}
\item $ t=0 $ is a branching point of the $ t^{\epsilon-1} $,
\item $ t^*_i=\frac{s_i}{S_3},\,i=2,5,6 $ are the branching point of the function $ g_i(t),\,i=2,5,6 $ (branching point of the square root),
\item $ t_{ai}=\frac{s_i}{S}\Bigl(1-\frac{r_{k(i)}^2}{\Delta}\Bigr) $, where the argument of the first function in~\eqref{eq:newrep} becomes $ -1 $,
\item $ t_{bi} $ and $ t_{ci} $: $ g_i^2(t_{bi,ci})=-\Delta\Bigl(\frac{S}{s_i}t_{bi,ci}-1\Bigr) $, where the argument of the second function in~\eqref{eq:newrep} becomes $ -1 $ (points $ t_{bi} $ and $ t_{ci} $ for $ i=2,5,6 $ can be in different sheets of function $ g_i(t),\,i=2,5,6 $ thanks of the square root), 
\item $ t_{0i} $: $ g_i(t_{i0})=0 $, where the argument of the second function in~\eqref{eq:newrep} becomes $ 0 $,
\item $ t_{\infty i}=\frac{s_i}{S} $, where the arguments of both functions in~\eqref{eq:newrep} become $ \infty $.
\end{itemize}
We will carry out an analytic continuation from the region $ {\cal R} $~\eqref{eq:region} to the region of interest by the paths lying in the region $ {\cal D}=\{\boldsymbol{s}|\Im s_i\geqslant0\} $ as in~\cite{KL:2016}.

\subsection*{Analytical continuation in Euclidean region}
Let us now discuss the analytical continuation of the result obtained in the region~\eqref{eq:region} to the whole Euclidean region. 

\begin{figure}
\begin{center}
\includegraphics[width=10cm]{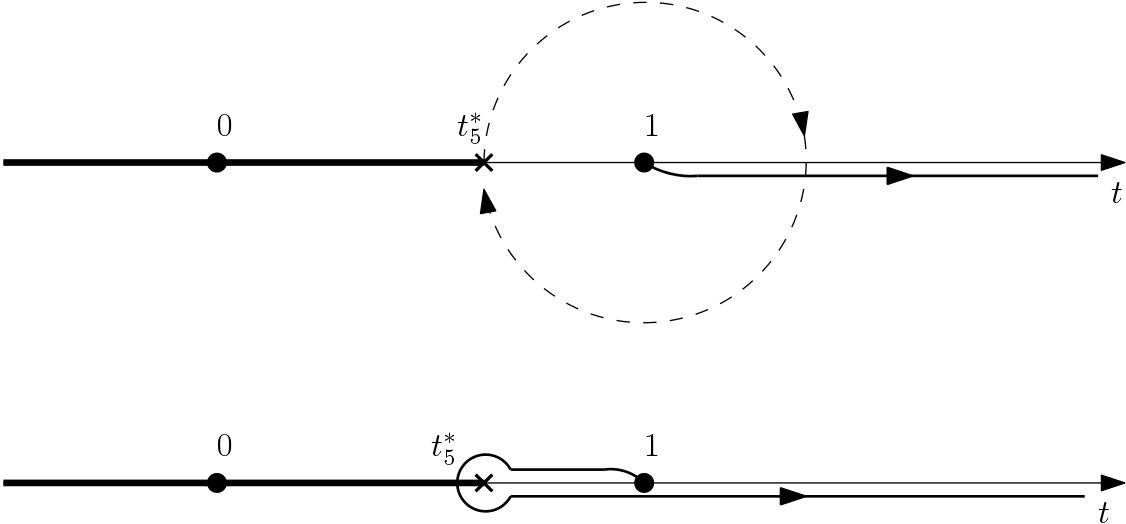}
\end{center}
\caption{The figure shows the movement of the singular point $ t_5^* $ and the transformation of integration contour.}\label{fig:2}
\end{figure}

Consider the analytical continuation of $ \widehat{P}_5 $ integral from region $x_2>0$, $x_6>0 $ and $ x_5<0 $ to $ x_5>0 $. We put $ s_5-s_2-\mu_1=re^{i\phi} $ and change $ \phi $ form $ \pi $ to $ 0 $ taking $ r $ sufficiently small so that the singular point $ t_5^* $ was the closest singular point to $ t=1 $. Then, only one singular point $ t_5^*$ is moving around point $ t=1 $, see Fig~\ref{fig:2}:
\begin{equation}
(s_5-s_2-\mu_1)^2+4s_2\mu_1(t_5^*-1)=0\,,\quad t_5^*=1-\frac{r^2}{4s_2\mu_1}e^{2i\phi}\,.
\end{equation}
From Fig.~\ref{fig:2} you can see how the path of integration changes, hence our integral transforms to
\begin{equation}\label{eq:p5x5}
\widehat{P}_5\rightarrow\widehat{P}_5+(-s_5)^{-\epsilon}\int_{t_5^*}^{\infty}\frac{dt\,t^{\epsilon-1}}{b_5(t)}\biggl[\arctan\frac{g_5(t)}{b_5(t)}-\arctan\frac{\bar{g}_5(t)}{b_5(t)}\biggr]\,,
\end{equation}
where $ \bar{g}_5(t) $ has a different sign before the square root in contrast to $ g_5(t) $
\begin{equation}
\bar{g}_5(t)=r_3+\frac{(s_3s_4-s_1\mu_1)(s_5-s_2-\mu_1)}{\mu_1}\biggl[1+\sqrt{1+\frac{4s_2\mu_1}{(s_5-s_2-\mu_1)^2}(t-1)}\biggr]\,.
\end{equation}
The additional integral in~\eqref{eq:p5x5} we can rewrite in the following form
\begin{equation}\label{eq:G}
\begin{split}
&(-s_5)^{-\epsilon}\int_{t_5^*}^{\infty}\frac{dt\,t^{\epsilon-1}}{b_5(t)}\biggl[\arctan\frac{g_5(t)}{b_5(t)}-\arctan\frac{\bar{g}_5(t)}{b_5(t)}\biggr]=\\
&=(-S_3)^{-\epsilon}\int_1^{\infty}\frac{dt\,t^{\epsilon-1}}{b(t)}\biggl[\arctan\frac{g_-(t)}{b(t)}-\arctan\frac{g_+(t)}{b(t)}\biggr]-{\cal H}(\boldsymbol{s},\epsilon)=G(\boldsymbol{s},\epsilon)-{\cal H}(\boldsymbol{s},\epsilon)
\end{split}
\end{equation}
where we introduce denotations
\begin{equation}
b(t)=\sqrt{\Delta\Bigl(\frac{S}{S_3}t-1\Bigr)}\;,\quad g_{\pm}=-s_2s_3+s_3s_4-s_4s_5\pm s_1\sqrt{\Delta_3(t-1)}\,.
\end{equation}
We can generalize the considered case to  analytic continuation for any sign of $ x_2,\,x_5,\,x_6 $ in the Euclidean region. In transition from a region of one sign $ x_i $ to the region with the other sign of $ x_i $, the singular point $ t^*_i=s_i/S_3 $ completely bypasses the point $ t = 1 $ meshing with the contour of integration as shown in Figure~\ref{fig:2}. Hence we can write our solution for any sign of $ \sigma_i=\sign(x_i)$ for Euclidean region:
\begin{equation}
\begin{split}
&P=\frac{2C(\epsilon)}{\epsilon}\biggl\{\sum_{i=1}^6\widehat{P}_i+\bigl(\Theta(\sigma_2)-\Theta(-\sigma_5)-\Theta(-\sigma_6)\bigr)G(\boldsymbol{s},\epsilon)+\\
&+{\cal H}(\boldsymbol{s},\epsilon)\Bigl(\Theta(-\sigma_2)-\Theta(\sigma_5)-\bigl[\Theta(\mu_1-s_3)+\Theta(\mu_1-s_4)-1\bigr]\Bigr)\biggr\}\,.
\end{split}
\end{equation}

Let us consider analytical continuation to the region with $ s_3s_4-s_1\mu_1<0  $ and $ \Delta>0 $. We start from region $ {\cal R} $, $ \mu_1>s_3 $, $ \mu_1>s_4 $ and go to $ s_3s_4-s_1\mu_1<0 $ by path $ s_1=\frac{s_3s_4}{\mu_1}-re^{i\phi} $, where $ \phi  $ changes from $ 0  $ to $ -\pi $. Hence we have
\begin{equation}\label{eq:transform2}
\begin{split}
&\widehat{P}_1\rightarrow\widehat{P}_1+\frac{i\pi}{2}(-s_1)^{-\epsilon}\int_{t_{a1}}^{+\infty}\frac{dt\,(t-i0)^{\epsilon-1}}{\sqrt{\Delta\Bigl(1-\frac{S}{s_1}t\Bigr)}}\,,\;\widehat{P}_3\rightarrow\widehat{P}_3-\frac{i\pi}{2}(-s_3)^{-\epsilon}\int_{t_{b3}}^{+\infty}\frac{dt\,(t-i0)^{\epsilon-1}}{\sqrt{\Delta\Bigl(1-\frac{S}{s_3}t\Bigr)}}\,,\\
&\widehat{P}_4\rightarrow\widehat{P}_4-\frac{i\pi}{2}(-s_4)^{-\epsilon}\int_{t_{c4}}^{+\infty}\frac{dt\,(t-i0)^{\epsilon-1}}{\sqrt{\Delta\Bigl(1-\frac{S}{s_4}t\Bigr)}}\,,\quad\widehat{P}_2\rightarrow\widehat{P}_2\,,\;\widehat{P}_5\rightarrow\widehat{P}_5\,,\\
&\widehat{P}_6\rightarrow\widehat{P}_6+i\pi(-\mu_1)^{-\epsilon}\int_{t_{\infty6}}^{\infty}\frac{dt\,(t-i0)^{\epsilon-1}}{\sqrt{\Delta\Bigl(1-\frac{S}{\mu_1}t\Bigr)}}-i\frac{\pi}{2}(-\mu_1)^{-\epsilon}\int_{+\infty}^{t_{a6}}\frac{dt\;t^{\epsilon-1}}{\sqrt{\Delta\Bigl(1-\frac{S}{\mu_1}t\Bigr)}}\,.
\end{split}
\end{equation}
Using relations
\begin{equation}
\begin{split}
&\frac{t_{a6}}{\mu_1}=\frac{t_{a1}}{s_1}=\frac{t_{b3}}{s_3}=\frac{t_{c4}}{s_4}\,,\\
&i\pi(-s_i)^{-\epsilon}\int_{t_{\infty i}}^{\infty}\frac{(t+i0)^{\epsilon-1}dt}{\sqrt{\Delta\Bigl(1-\frac{S}{s_i}t\Bigr)}}=\pi^{\frac{3}{2}}e^{i\pi\epsilon}\frac{S^{-\epsilon}}{\sqrt{\Delta}}\frac{\Gamma(1/2-\epsilon)}{\Gamma(1-\epsilon)}={\cal H}(\bm{s},\epsilon)\,,
\end{split}
\end{equation}
the sum of $ \widehat{P}_i $ in~\eqref{eq:transform2} is transformed to
\begin{equation}
\begin{split}
&\sum_{i=1}^6\widehat{P}_i\rightarrow\sum_{i=1}^6\widehat{P}_i+{\cal H}(\boldsymbol{s},\epsilon)\,,\quad\widehat{P}\rightarrow\sum_{i=1}^6\widehat{P}_i\,.
\end{split}
\end{equation}
Therefore all terms with the homogenius solution vanish. For other subregions of Euclidean region analytical continuation is performed in the same way. Hence we can write the solution for the whole Euclidean region
\begin{equation}
\begin{split}
&P^{(6-2\epsilon)}=\frac{2C(\epsilon)}{\epsilon}\biggl(\sum_{i=1}^6\widehat{P}_i+\Bigl(\Theta(\sigma_5)-\Theta(-\sigma_2)-\Theta(\sigma_6)\Bigr)G(\boldsymbol{s},\epsilon)+\\
&+\Theta(s_3s_4-s_1\mu_1)\Bigl(\Theta(-\sigma_2)-\Theta(\sigma_5)-\Theta(\mu_1-s_3)-\Theta(\mu_1-s_4)+1\Bigr){\cal H}(\boldsymbol{s},\epsilon)\biggr)\,.
\end{split}
\end{equation}

\subsection*{Analytical continuation to other regions}

Here we consider analytical continuation from Euclidean region to regions with positive sign of invariants. We have $ 2^6 $ regions of invariants with sign $ + $ or $ - $, but we have symmetry of the integral and the identity 
\begin{equation}
P(s_1,s_2,s_3,s_4,s_5,\mu_1)=P(s_1,s_5,s_4,s_3,s_2,\mu_1)\,,\quad P^{(6-2\epsilon)}(\boldsymbol{s})=e^{i\pi\epsilon}\Bigl[P^{(6-2\epsilon)}(-\boldsymbol{s})\Bigr]^*\,.
\end{equation}
Therefore there are $ 20 $ non-equivalent regions of invariants (see table~\ref{tab:regions}). In the table we use the denotaion for regions $ \bigl(\sign(s_1),\sign(s_2),\sign(s_3),\sign(s_4),\sign(s_5),\sign(\mu_1)\bigr) $.
\begin{table}
\begin{center}
\begin{tabular}{cccc}
0 plus & 1 plus & 2 plus & 3 plus\\
\hline
$ {\cal R}_1\;(------) $ & 
\begin{tabular}{l}
${\cal R}_2\; (+-----) $\\ ${\cal R}_3\; (-+----) $\\ ${\cal R}_4\; (--+---) $\\ ${\cal R}_5\; (-----+) $
\end{tabular}
&
\begin{tabular}{l}
${\cal R}_6\; (++----) $\\ ${\cal R}_7\; (+-+---) $\\ ${\cal R}_8\; (+----+) $\\ ${\cal R}_9\; (-+---+) $\\ ${\cal R}_{10}\; (--+--+) $\\ ${\cal R}_{11}\; (-++---) $\\ ${\cal R}_{12}\;  (--++--) $\\ ${\cal R}_{13}\;  (-+-+--) $\\ ${\cal R}_{14}\;  (-+--+-) $
\end{tabular}
&
\begin{tabular}{l}
${\cal R}_{15}\;  (+++---) $\\ ${\cal R}_{16}\;  (++-+--) $\\ ${\cal R}_{17}\;  (++--+-) $\\ ${\cal R}_{18}\;  (++---+) $\\ ${\cal R}_{19}\;  (+-++--) $\\ ${\cal R}_{20}\;  (+-+--+) $
\end{tabular}\\
\hline
\end{tabular}
\caption{20 non-equivalent regions of invariants.}\label{tab:regions}
\end{center}
\end{table}

Let us consider the analytic continuation into the most interesting region with $ s_2>0 $. We start from Euclidean region with $ x_2>0,\,x_5<0,\,x_6>0 $, $ \Delta>0 $ and $ s_3s_4-s_1\mu_1<0 $, where there is no homogeneous solution. We put $ s_2=|s_2|e^{i\phi} $ and change $ \phi $ from $ \pi $ to $ 0 $. While changing $ \phi $, we track the motion of the singular points $ t_{ai},\,t_{bi},\,t_{ci},\,t_{0i},\,t_{\infty i},\,t^*_i $ and deform the integration contours over $ t $ in such a way that they do not cross these singular points (and also $ t=0 $ ). In final position, when $ \phi=0 $ the integrals are written as
\begin{equation}\label{eq:ac-s2}
\begin{split}
&\widehat{P}_1\rightarrow\widehat{P}_1+\frac{\pi}{2}\frac{(-s_1)^{-\epsilon}}{\sqrt{\Delta}}\biggl(i\int_{t_{b1}}^{t_{\infty1}}\frac{dt\,t^{\epsilon-1}}{\sqrt{1-S/s_1 t}}+\int_{t_{\infty1}}^{\infty}\frac{dt\,t^{\epsilon-1}}{\sqrt{S/s_1 t-1}}\biggr)\,,\\
&\widehat{P}_2\rightarrow\widehat{P}_2\,,\\
&\widehat{P}_3\rightarrow\widehat{P}_3+\frac{\pi}{2}\frac{(-s_3)^{-\epsilon}}{\sqrt{\Delta}}\biggl(-i\int_{t_{b3}}^{t_{\infty3}}\frac{dt\,t^{\epsilon-1}}{\sqrt{1-S/s_3 t}}-\int_{t_{\infty3}}^{\infty}\frac{dt\,t^{\epsilon-1}}{\sqrt{S/s_3 t-1}}\biggr)\,,\\
&\widehat{P}_4\rightarrow\widehat{P}_4+\frac{\pi}{2}\frac{(-s_4)^{-\epsilon}}{\sqrt{\Delta}}\biggl(-i\int_{t_{a4}}^{t_{\infty4}}\frac{dt\,t^{\epsilon-1}}{\sqrt{1-S/s_4 t}}+\int_{t_{\infty 4} }^{\infty}\frac{dt\,t^{\epsilon-1}}{\sqrt{S/s_4 t-1}}\biggr)\,,\\
&\widehat{P}_5\rightarrow\widehat{P}_5+\frac{i\pi}{2}\frac{(-s_5)^{-\epsilon}}{\sqrt{\Delta}}\int_{t_{b5}}^{t_{5\infty}}\frac{dt\,t^{\epsilon-1}}{\sqrt{1-S/s_5 t}}-\frac{1}{2}G(\boldsymbol{s},\epsilon)\,,\\
&\widehat{P}_6\rightarrow\widehat{P}_6-\frac{\pi}{2}\frac{(-\mu_1)^{-\epsilon}}{\sqrt{\Delta}}\int_{t_{\infty 6}}^{\infty}\frac{dt\,t^{\epsilon-1}}{\sqrt{S/\mu_1 t-1}}+\frac{1}{2}G(\boldsymbol{s},\epsilon)\,.
\end{split}
\end{equation}
Using the relations
\begin{equation}
\begin{split}
&\frac{t_{a1}}{s_1}=\frac{t_{c3}}{s_3}=\frac{t_{b4}}{s_4}=\frac{t_{a6}}{\mu_1}=\frac{s_3+s_4-s_1-\mu_1}{s_3s_4-s_1\mu_1}\,;\quad
\frac{t_{b1}}{s_1}=\frac{t_{a2}}{s_2}=\frac{t_{a4}}{s_4}=\frac{s_1+s_2-s_4}{s_1s_5}\,;\\
&\frac{t_{c1}}{s_1}=\frac{t_{a3}}{s_3}=\frac{t_{a5}}{s_5}=\frac{s_1+s_5-s_3}{s_1s_5}\,;
\quad\frac{t_{2b}}{s_2}=\frac{t_{3b}}{s_3}=\frac{t_{5b}}{s_5}=\frac{t_{6b}}{\mu_1}=\frac{s_2s_3+(\mu_1-s_3)(s_5-s_3)}{s_3^2s_2}\,;\\
&\frac{t_{2c}}{s_2}=\frac{t_{4c}}{s_4}=\frac{t_{5c}}{s_5}=\frac{t_{6c}}{\mu_1}=\frac{s_4s_5+(\mu_1-s_4)(s_2-s_4)}{s_4^2s_5}\,,
\end{split}
\end{equation}
the sum of the terms in~\eqref{eq:ac-s2} is transformed to $ \sum_{i=1}^6\widehat{P}_i $ without the homogeneous solution. Analytical continuation in other regions can be performed in a similar way. Since there are a lot of regions, it is a difficult task. To simplify the analytical continuation it is possible to use the fact that the pentagon must be finite for $ d>4 $ as $ \epsilon\rightarrow0$. Then we look for a solution in an arbitrary domain in the form of
\begin{equation}\label{eq:hyp1}
\widehat{P}=\sum_{i=1}^6\widehat{P}_i+aG(\boldsymbol{s},\epsilon)+b\Theta(\Delta){\cal H}(\boldsymbol{s},\epsilon)\,,
\end{equation}
where coefficients $a,\,b$ are integers. One can pick coefficients $a$ and $b$ with the help of numerical calculation of the formula \eqref{eq:hyp1} and demanding that the result should be of order $\epsilon^0$. In each non-equivalent region (see table~\ref{tab:1} or~\ref{tab:regions}) there are subregions  defined by the following thresholds (signs of the following expressions) 
\begin{itemize}
\item $ \sign(s_3s_4-s_1\mu_1) $ affects  position of $ t_{\infty i} $ singular points,
\item $ \sign(s_3-\mu_1) $ and $ \sign(s_4-\mu_1) $ affect the location of the singular points in the $ \widehat{P}_3 $ and $ \widehat{P}_4 $,
\item $ \sign(s_2-s_5-\mu_1) $, $ \sign(s_5-s_2-\mu_1) $ and $ \sign(\mu_1-s_2-s_5) $ affect the location of $ t_i^* $ singular points,
\item $ S-S_3>0 $ together with $ \sign(s_2s_3-s_3s_4+s_4s_5) $ and $ \sign(s_3s_4(s_5-s_2)/\mu_1+s_2s_3-s_4s_5) $  determine the possibility that the  integration contour crosses the cut from singular point $ t_{\infty i} $.
\end{itemize}
We have to pick the coefficients $ a $ and $ b $  for each subregion of region $ {\cal R}_i $. The result for analytical continuation into the different regions of invariants can be found in the table~\ref{tab:1}.

\section{Conclusion}

In this work we have calculated the pentagon integral using a system of differential equations that was solved by bringing it to the $ \epsilon $-form. Solution for the integral is a sum of one fold integrals. In dimension $ d>4 $ the pentagon integral is finite and trivial for $ \epsilon $ expansion.

\begin{acknowledgments}
This work is supported by the RFBR grants No. 16-32-60033 and 16-02-00888.
\end{acknowledgments}


\appendix

\section{Triangle integral}\label{sec:tri}
\begin{figure}
\begin{center}
\includegraphics[width=5cm]{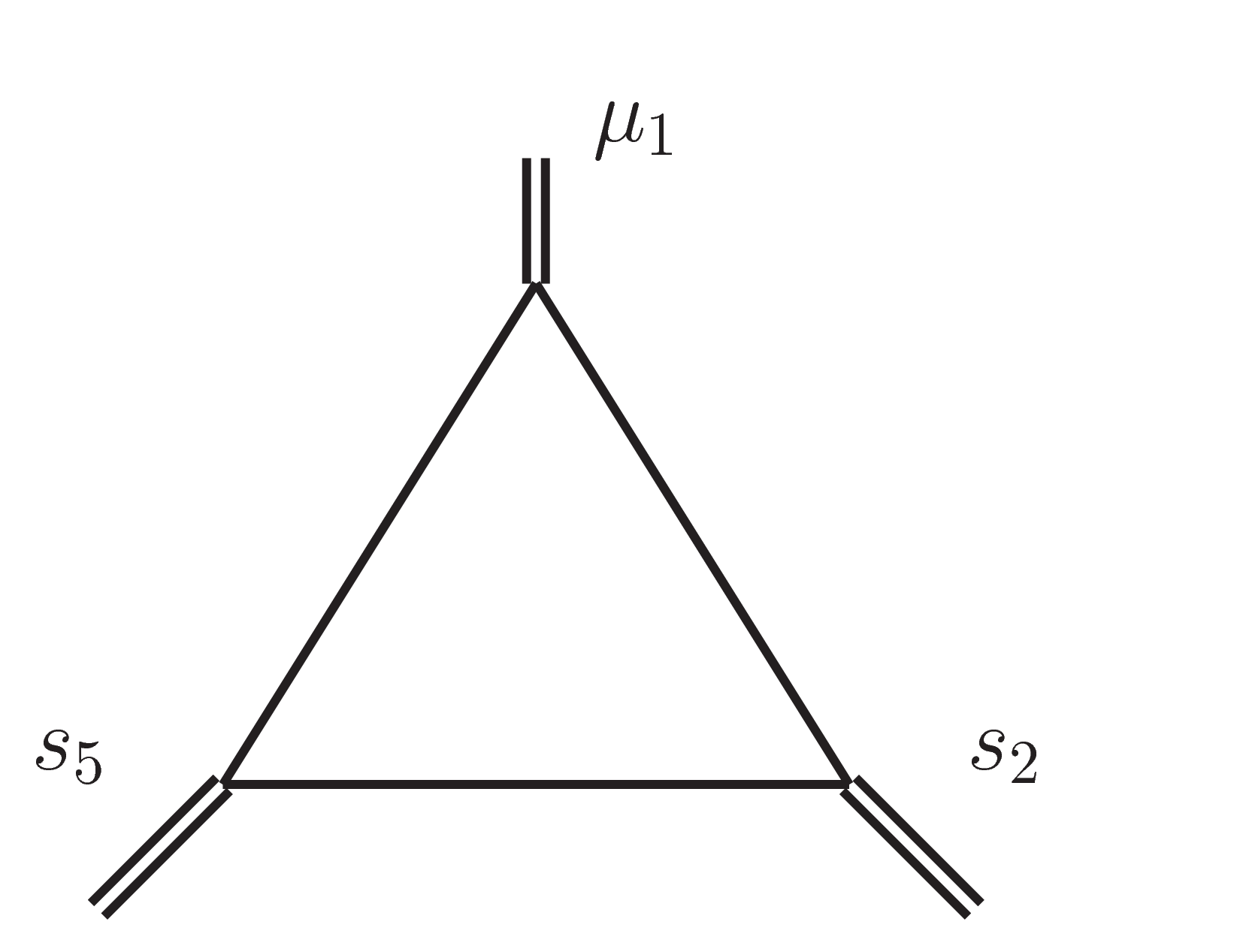}
\end{center}
\caption{Schematic representation of the Triangle master integral}\label{fig:tri}
\end{figure}
We use IBP reduction to obtain the system of partial differential equations for the triangle integral $ T $ and three simpler master integrals $ R_2,\,R_5,\,R_6 $, where
\begin{equation}
R_{2,5}=R(s_{2,5})\;,\quad R_6=R(\mu_1)\,.
\end{equation}
For the triangle integral we introduce the column-vector
\[
\boldsymbol{J}_T=(T,R_2,R_5,R_6)^T\,.
\]
Next, we find an appropriate basis to reduce the system to $ \epsilon $-form
\begin{equation}
T=\frac{C(\epsilon)}{\epsilon^2\sqrt{\Delta_3}}\widetilde{T}\,,\quad R_i=\frac{C(\epsilon)}{\epsilon(1-2\epsilon)}\widetilde{R}_i\,.
\end{equation}
The differential equations in the new basis can be written in $ d\log $-form
\begin{equation}
\begin{split}
&d\widetilde{T}=-\epsilon\widetilde{T}d\log\Bigl(-S_3\Bigr)-2\epsilon \widetilde{R}_2d\arctan x_2-2\epsilon \widetilde{R}_5d\arctan x_5-2\epsilon \widetilde{R}_6d\arctan x_6\,,\\
&d\widetilde{R}_i=-\epsilon\widetilde{R}_id\log s_i\;,\quad i=2,5\,,\;d\widetilde{R}_6=-\epsilon\widetilde{R}_6d\log \mu_1\;.
\end{split}
\end{equation}
where
\begin{equation}
x_2=\frac{s_2-s_5-\mu_1}{\sqrt{\Delta_3}}\,,\;x_5=\frac{s_5-s_2-\mu_1}{\sqrt{\Delta_3}}\,,\;x_6=\frac{\mu_1-s_2-s_5}{\sqrt{\Delta_3}}\,.
\end{equation}
We search solution in the form $ \widetilde{J}_T=\widetilde{J}_T^{(1)}+\widetilde{J}_T^{(2)}+\widetilde{J}_T^{(3)} $, where
\begin{equation}
\widetilde{J}_T^{(1)}=(\widetilde{T}^{(2)},\widetilde{R}_2,0,0)^T\,,\;\widetilde{J}_T^{(2)}=(\widetilde{T}^{(5)},0,\widetilde{R}_5,0)^T\,,\;\widetilde{J}_T^{(3)}=(\widetilde{T}^{(6)},0,0\widetilde{R}_6)^,.
\end{equation}
Therefore we arrive at the following differential equation for $ \widetilde{T}_1 $ (the other equations are the same):
\begin{equation}\label{eq:trieq}
d\Bigl((-S_3)^{\epsilon}\widetilde{T}^{(2)}\Bigr)=-2\epsilon \bigl(S_3/s_2\bigr)^{\epsilon} d\arctan x_2\,.
\end{equation}
The right-hand side of Eq.~\eqref{eq:trieq} depends only on $ x_2 $. In particular 
\begin{equation}
S_3=s_2(1+x_2^2)=s_5(1+x_5^2)=\mu_1(1+x_6^2)\,.
\end{equation}
Let us consider the region of invariants $ x_2>0,\,x_5<0,\,x_6>0,\,\Delta_3>0 $. Then from the differential equations we have
\begin{equation}\label{eq:T}
\begin{split}
&\widetilde{T}^{(2)}=-2\epsilon(-S_3)^{-\epsilon}\Re\int_{+\infty}^{x_2}(1+z^2)^{\epsilon-1}dz\,,\quad\widetilde{T}^{(5)}=-2\epsilon(-S_3)^{-\epsilon}\Re\int_{-\infty}^{x_5}(1+z^2)^{\epsilon-1}dz\\
&\widetilde{T}^{(6)}=-2\epsilon(-S_3)^{-\epsilon}\Re\int_{+\infty}^{x_6}(1+z^2)^{\epsilon-1}dz
\end{split}
\end{equation}
Our solution has the following form
\begin{equation}
T=\frac{C(\epsilon)}{\epsilon^2\sqrt{\Delta_3}}\biggl(\widetilde{T}^{(2)}+\widetilde{T}^{(5)}+\widetilde{T}^{(6)}+g(\epsilon)(-S_3)^{-\epsilon}\biggr)\,,
\end{equation}
where $ g(\epsilon) $ is a constant of integration. In the limit $ \Delta_3\rightarrow0 $ the triangle integral can be expressed as a linear combination of three bubble integrals
\begin{equation}
T\Big|_{\Delta_3\rightarrow0}\approx-\frac{C(\epsilon)}{\epsilon(1-2\epsilon)}\biggl(\frac{(-s_2)^{-\epsilon}}{s_2-s_5-\mu_1}-\frac{(-s_5)^{-\epsilon}}{s_5-s_2-\mu_1}+\frac{(-\mu_1)^{-\epsilon}}{\mu_1-s_2-s_5}\biggr)+\frac{C(\epsilon)g(\epsilon)}{\epsilon^2\sqrt{\Delta_3}}\Bigl(-\frac{4s_2s_5\mu_1}{\Delta_3}\Bigr)^{-\epsilon}\,.
\end{equation}
Therefore $ g(\epsilon)=0 $ and for $ x_2>0,\,x_5<0,\,x_6>0 $
\begin{equation}
T=\frac{C(\epsilon)}{\epsilon^2\sqrt{\Delta_3} }\biggl(\widetilde{T}^{(2)}+\widetilde{T}^{(5)}+\widetilde{T}^{(6)}\biggr)\,.
\end{equation}


\section{Easy box}\label{sec:eb}
\begin{figure}
\begin{center}
\includegraphics[width=6cm]{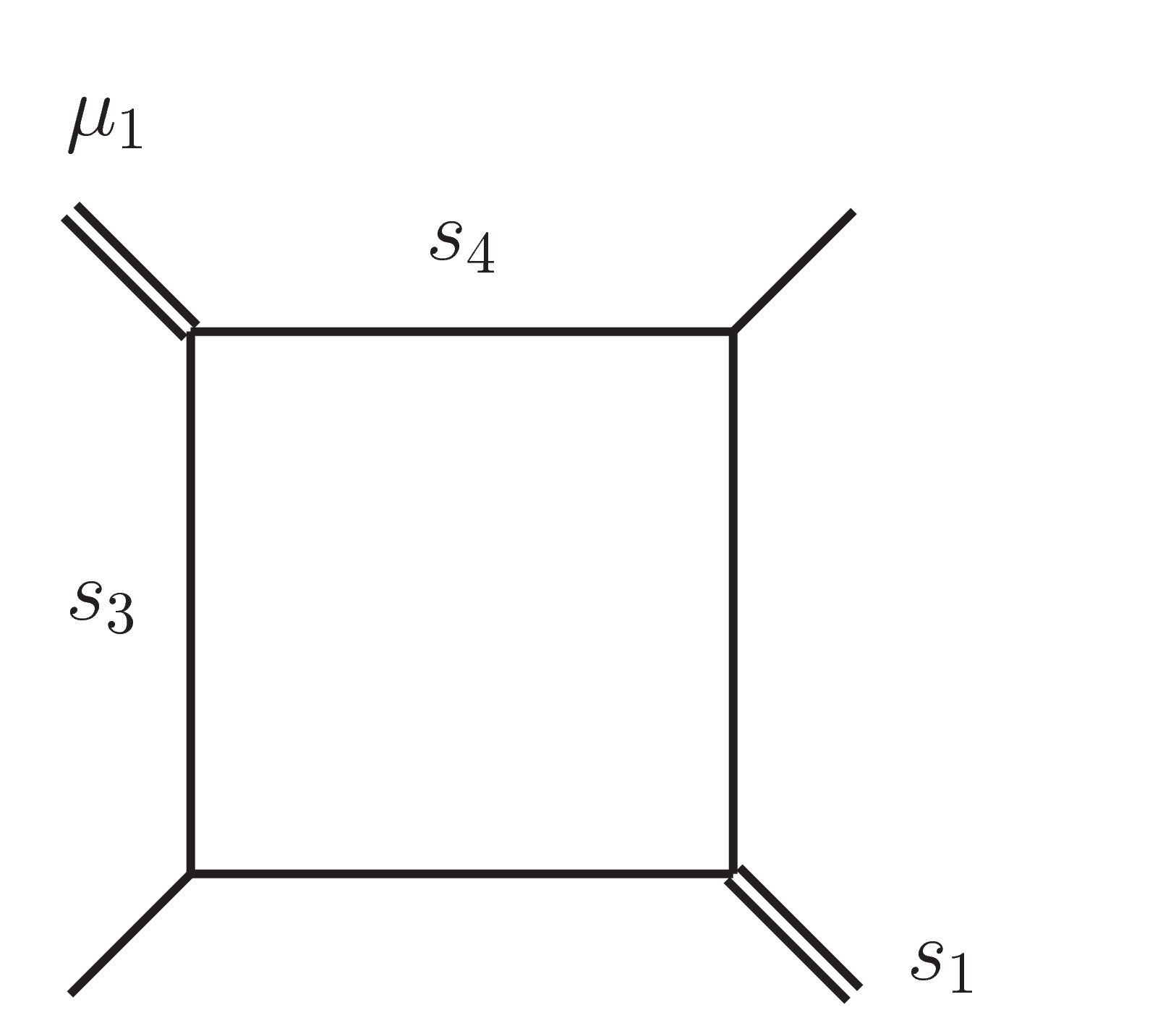}
\end{center}
\caption{Schematic representation of the easy box master integral}\label{fig:eb}
\end{figure}
In this section we consider Easy box master integral in $ d=4-2\epsilon $ dimensions in ``Euclidean'' region $ s_1<0,\,\mu_1<0,\,s_3<0,\,s_4<0 $ (see Fig.~\ref{fig:eb}). For the easy box integral we introduce  the column vector
\begin{equation}
\boldsymbol{J}_E=\bigl(B_1,R(s_1),R(\mu_1),R(s_3),R(s_4)\bigr)^T\,.
\end{equation}
Next, we find an appropriate basis to reduce the system to $ \epsilon $-form
\begin{equation}
B_1=\frac{C(\epsilon)}{\epsilon^2(s_3s_4-s_1\mu_1)}\widetilde{B}_1\;,\quad R(s_i)=\frac{C(\epsilon)}{\epsilon(1-2\epsilon )}\widetilde{R}(s_i)\,,\;\widetilde{R}(s_i)=(-s_i)^{-\epsilon}\,.
\end{equation}
The differential equations in the new basis have $ \epsilon $-form and can be written in $ d\log $-form
\begin{equation}
\begin{split}
d\widetilde{B}_1&=-\epsilon\biggl\{\widetilde{B}_1d\log(-S_{4e})-2\widetilde{R}(s_1)d\log(y_1-1)-2\widetilde{R}(\mu_1)d\log(y_2-1)+\\
&+2\widetilde{R}(s_3)d\log(y_3-1)+2\widetilde{R}(s_4)d\log(y_4-1)\biggr\}\,,
\end{split}
\end{equation}
where
\begin{equation}
(-S_{4e})=\frac{s_3s_4-s_1\mu_1}{s_3+s_4-s_1-\mu_1}=(-s_1)y_1=(-\mu_1)y_2=(-s_3)y_3=(-s_4)y_4\,.
\end{equation}
We search solution in the form $ \widetilde{J}_{4e}=\sum_{i=1}^4\widetilde{J}^{(i)}$ where
\begin{equation}
\begin{split}
&\widetilde{J}^{(1)}=(\widetilde{B}_1^{(1)},\widetilde{R}(s_1),0,0,0),\;\widetilde{J}^{(2)}=(\widetilde{B}_1^{(2)},0,\widetilde{R}(\mu_1),0,0),\;\\
&\widetilde{J}^{(3)}=(\widetilde{B}_1^{(3)},0,0,\widetilde{R}(s_3),0),\;\widetilde{J}^{(4)}=(\widetilde{B}_1^{(4)},0,0,0,\widetilde{R}(s_4)),\;
\end{split}
\end{equation}
hence for $ \widetilde{B}_1^{(1)} $  we have the following equation
\begin{equation}
d\Bigl((-S_{4e})^{\epsilon}\widetilde{B}_1^{(1)}\Bigr)=2\epsilon \Bigl(\frac{S_{4e}}{s_1}\Bigr)^{\epsilon}d\log(y_1-1)=2\epsilon y_1^{\epsilon}d\log(y_1-1)\;.
\end{equation}
Then from the differential equation we have
\begin{equation}
\begin{split}
&\widetilde{B}_1^{(1)}=2\epsilon\, (-S_{4e})^{-\epsilon}\Re\int_{\infty}^{y_1}\frac{y^{\epsilon}dy}{y-1+i0}=-2\epsilon(-s_1)^{-\epsilon}\Re\int^{\infty}_{1}\frac{z^{\epsilon}dz}{z-s_1/S_{4e}+i0}=\\
&=2(-s_1)^{-\epsilon}\Re\,_2F_1\Bigl(1,-\epsilon;1-\epsilon;\frac{s_1}{S_{4e}}\Bigr)\,,
\end{split}
\end{equation}
\begin{equation}
\begin{split}
&\widetilde{B}_1^{(1)}=2(-s_1)^{-\epsilon}\Re\,_2F_1\Bigl(1,-\epsilon;1-\epsilon;\frac{s_1}{S_{4e}}\Bigr)\,,\;\widetilde{B}_1^{(2)}=2(-\mu_1)^{-\epsilon}\Re\,_2F_1\Bigl(1,-\epsilon;1-\epsilon;\frac{\mu_1}{S_{4e}}\Bigr)\,,\\
&\widetilde{B}_1^{(3,4)}=-2(-s_{3,4})^{-\epsilon}\Re\,_2F_1\Bigl(1,-\epsilon;1-\epsilon;\frac{s_{3,4}}{S_{4e}}\Bigr)\,.
\end{split}
\end{equation}
We can write the solution for the easy box integral in the form
\begin{equation}
B_1=\frac{C(\epsilon)}{\epsilon^2(s_3s_4-s_1\mu_1)}\biggl(\sum_{i=1}^4\widetilde{B}_1^{(i)}+g(\epsilon)(-S_{4e})^{-\epsilon}\biggr)\,.
\end{equation}
Let us consider Gramm determinant for the easy box
\begin{equation}
\Delta_{4e}=\det\Bigl(2(p_i,p_j)\Bigr)=-2(s_3+s_4-s_1-\mu_1)(s_3s_4-s_1\mu_1)\,.
\end{equation}
Therefore in the limit $ \Delta_{4e}\rightarrow 0\;(S_{4e}\rightarrow0) $ we have
\begin{equation}
\widetilde{B}_1^{(1)}\rightarrow 2\,(-S_{4e})^{-\epsilon} \Gamma(1-\epsilon)\Gamma(1+\epsilon)\,.
\end{equation}
Since $ \Delta_{4e}=0 $ is not a branching point of $ B_1 $ hence $ g(\epsilon)=0 $.

Solution for the easy box has the form
\begin{equation}
\begin{split}
&B_1=\frac{2C(\epsilon)}{\epsilon^2(s_3s_4-s_1\mu_1)}\Biggl\{(-s_1)^{-\epsilon}\Re\,_2F_1\Bigl(1,-\epsilon;1-\epsilon;\frac{s_1(s_1+\mu_1-s_3-s_4)}{s_3s_4-s_1\mu_1}\Bigr)+\\
&+(-\mu_1)^{-\epsilon}\Re\,_2F_1\Bigl(1,-\epsilon;1-\epsilon;\frac{\mu_1(s_1+\mu_1-s_3-s_4)}{s_3s_4-s_1\mu_1}\Bigr)-\\
&-(-s_3)^{-\epsilon}\Re\,_2F_1\Bigl(1,-\epsilon;1-\epsilon;\frac{s_3(s_1+\mu_1-s_3-s_4)}{s_3s_4-s_1\mu_1}\Bigr)-\\
&-(-s_4)^{-\epsilon}\Re\,_2F_1\Bigl(1,-\epsilon;1-\epsilon;\frac{s_4(s_1+\mu_1-s_3-s_4)}{s_3s_4-s_1\mu_1}\Bigr)\Biggr\}\,.
\end{split}
\end{equation}


\section{Hard box}\label{sec:hb}

Our hard box integral depends on four variables $ s,\,t,\,m_1^2=\mu_1,\,m_2^2=\mu_2 $ (see Fig.~\ref{fig:hb}). In this section we consider integrals in $ d=4-2\epsilon $ dimensions in ``Euclidean'' region $s<0,\,t<0,\,\mu_1<0,\,\mu_2<0 $. Let us introduce useful equalities
\begin{equation}
\Delta_{4h}=-2s\bigl((t-\mu_1)(t-\mu_2)+s\,t\bigr),\,S_{4h}=\frac{2s^2t^2}{\Delta_{4h}},\;\Delta_3=4\mu_1\mu_2-(s-\mu_1-\mu_2)^2,\,S_3=\frac{4s\mu_1\mu_2}{\Delta_3}\,.
\end{equation} 
We use IBP reduction to obtain the system of partial differential equations for the hard box integral $ H $ and five simpler master integrals. Introducing the column-vector
\begin{equation}
\boldsymbol{J}_H=\bigl(H,T,R(s),R(t),R(\mu_1),R(\mu_2)\bigr)^T\,,
\end{equation}
we may represent the system in the matrix form 
\begin{equation}
\frac{\partial\boldsymbol{J}_H}{\partial s_i}=M_i(\boldsymbol{s},\epsilon)\boldsymbol{J}_H\,,\;\boldsymbol{s}=(s,t,\mu_1,\mu_2)\,.
\end{equation}
We find an appropriate basis in order to reduce the system to $ \epsilon $-form:
\begin{figure}
\begin{center}
\includegraphics[width=5cm]{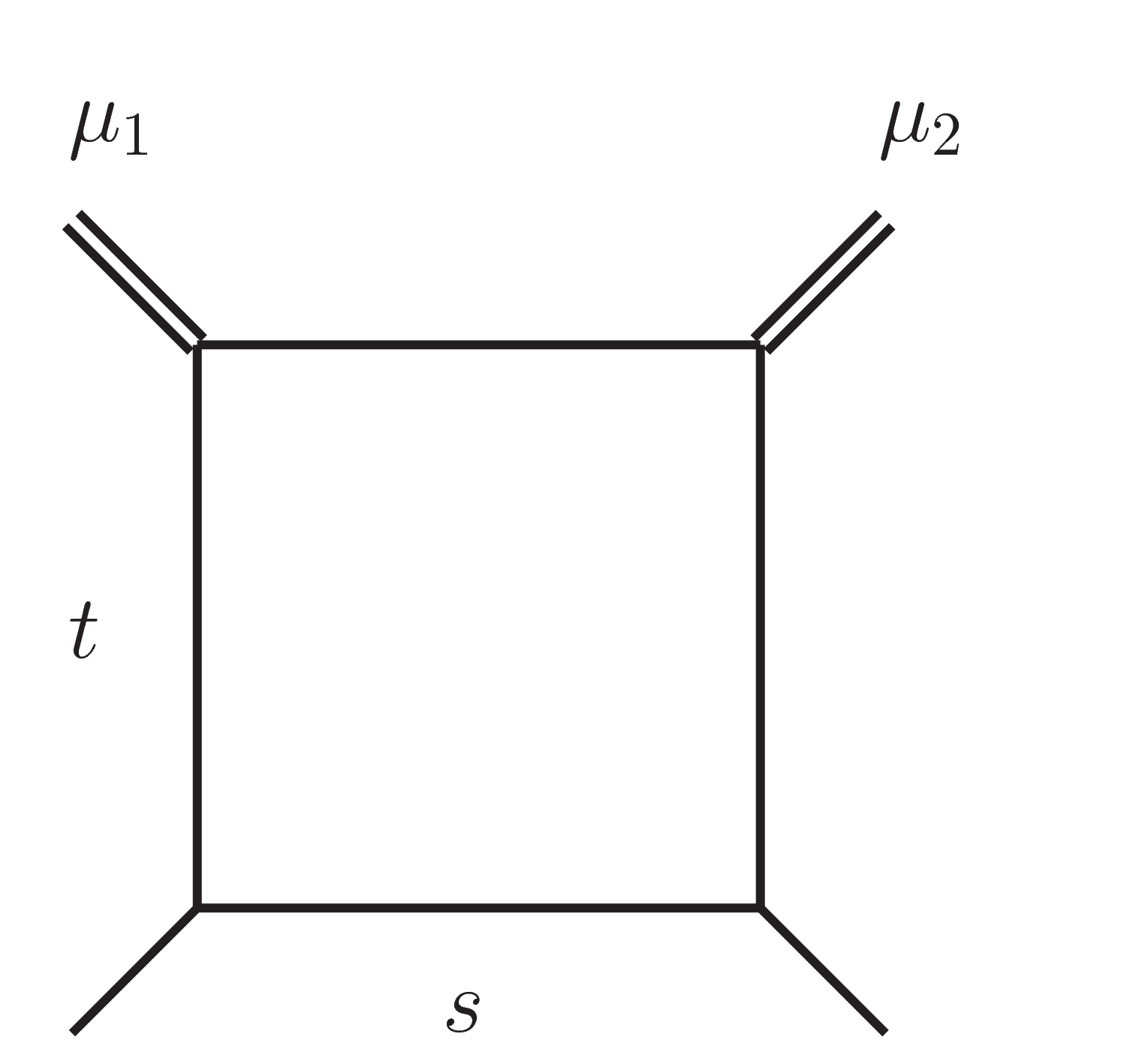}
\end{center}
\caption{Schematic representation of the hard box master integral}\label{fig:hb}
\end{figure}
\begin{equation}\label{eq:HBbasis}
H=\frac{C(\epsilon)}{\epsilon^2\,s\,t}\widetilde{H}\,,\quad T=\frac{C(\epsilon)}{\epsilon^2\sqrt{\Delta_3}}\widetilde{T}\,,\quad R(s_i)=\frac{C(\epsilon)}{\epsilon(1-2\epsilon)}\widetilde{R}(s_i)\,.
\end{equation}
The differential equation in the new basis can be written in $ d\log $-form
\begin{equation}
\begin{split}
&d\widetilde{H}=-\epsilon\Bigl(\widetilde{H}d\log(-S_{4h})-2\widetilde{T}d\arctan b+2(-t)^{-\epsilon}d\log(a-1)-\\
&-(-s)^{-\epsilon}d\log\frac{1+b^2}{(b-x_3)^2}+(-\mu_1)^{-\epsilon}d\log\frac{1+b^2}{(b+x_1)^2}+(-\mu_2)^{-\epsilon}d\log\frac{1+b^2}{(b+x_2)^2}\Bigr)\,,
\end{split}
\end{equation}
where we use denotations
\begin{equation}
\begin{split}
&x_1=\frac{\mu_1-\mu_2-s}{\sqrt{\Delta_3}}\,,\;x_2=\frac{\mu_2-\mu_1-s}{\sqrt{\Delta_3}}\,,\;x_3=\frac{s-\mu_1-\mu_2}{\sqrt{\Delta_3}}\,,\\
&b=\frac{t(s-\mu_1-\mu_2)+2\mu_1\mu_2}{t\sqrt{\Delta_3}}\,,\;a=\frac{s\,t}{(t-\mu_1)(t-\mu_2)+s\,t}\,,\\
&S_{4h}=(-t)a=(-s)\frac{1+x_3^2}{1+b^2}=(-\mu_1)\frac{1+x_1^2}{1+b^2}=(-\mu_2)\frac{1+x_2^2}{1+b^2}=(-S_3)\frac{1}{1+b^2}\,.
\end{split}
\end{equation}
We search for the solution in the form
\begin{equation}
\widetilde{\boldsymbol{J}}_H=\sum_{i=1}^4 \widetilde{J}^{(i)}\,,
\end{equation}
\begin{equation}
\begin{split}
&\widetilde{J}^{(3)}=\Bigl(\widetilde{H}^{(3)},\widetilde{T}^{(3)},\widetilde{R}(s),0,0,0\Bigr)\,,\;\widetilde{J}^{(1)}=\Bigl(\widetilde{H}^{(1)},\widetilde{T}^{(1)},0,0,\widetilde{R}(\mu_1),0\Bigr)\\
&\widetilde{J}^{(2)}=\Bigl(\widetilde{H}^{(2)},\widetilde{T}^{(2)},0,0,0,\widetilde{R}(\mu_2)\Bigr)\,,\;\widetilde{J}^{(4)}=\Bigl(\widetilde{H}^{(4)},0,0,\widetilde{R}(t),0,0\Bigr)\,.
\end{split}
\end{equation}
Let us consider the following region of invariants: $ \Delta_3>0,\,x_1>0,\,x_2<0,\,x_3>0 $. Therefore the triangle integral has form
\begin{equation}
\begin{split}
&\widetilde{T}=\sum_{i=1}^3\widetilde{T}^{(i)},\quad\widetilde{T}^{(3)}=-2\epsilon(-S_3)^{-\epsilon}\Re\int_{+\infty}^{x_3}(1+z^2)^{\epsilon-1}dz\,,\\
&\widetilde{T}^{(1)}=-2\epsilon(-S_3)^{-\epsilon}\Re\int_{+\infty}^{x_1}(1+z^2)^{\epsilon-1}dz,\;\widetilde{T}^{(2)}=-2\epsilon(-S_3)^{-\epsilon}\Re\int_{-\infty}^{x_2}(1+z^2)^{\epsilon-1}dz\,,
\end{split}
\end{equation}
where
\begin{equation}
S_3=\frac{4\mu_1\mu_2 s}{4\mu_1\mu_2-(s-\mu_1-\mu_2)^2}=(-s)(1+x_3^2)=(-\mu_1)(1+x_1^2)=(-\mu_2)(1+x_2^2)\,.
\end{equation}

Using the expression for the triangle integral we arrive at the following differential equations for $ \widetilde{H}^{(i)} $ 
\begin{equation}\label{eq:Hdif}
d\Bigl((-S_{4h})^{\epsilon}\widetilde{H}^{(i)}\Bigr)=G^{(i)}(x_i,b)db+F^{(i)}(x_i,b)dx_i\,,\quad i=1,2,3\,, \\
\end{equation}
\begin{equation}\label{eq:Hsimpl}
d\Bigl((-S_{4h})^{\epsilon}\widetilde{H}^{(4)}\Bigr)=-2\epsilon a^{\epsilon}d\log(a-1)\,,
\end{equation}
where
\begin{equation}
F^{(1,2)}(x_{1,2},b)=2\epsilon\biggl(\frac{1+x_{1,2}^2}{1+b^2}\biggr)^{\epsilon}\frac{1}{b-x_{1,2}}\,,\quad F^{(3)}(x_3,b)=-2\epsilon\biggl(\frac{1+x_3^2}{1+b^2}\biggr)^{\epsilon}\frac{1}{b+x_3}\,,
\end{equation}
\begin{equation}
\begin{split}
&G^{(1,2)}(x_{1,2},b)=2\epsilon(1+b^2)^{-1-\epsilon}(1+x_{1,2}^2)^{\epsilon}\biggl(-2\epsilon\int_{\pm\infty}^{x_{1,2}}\frac{(1+z^2)^{\epsilon-1}dz}{(1+x_{1,2}^2)^{\epsilon}}+b-\frac{1+b^2}{b-x_{1,2}}\biggr)\,,\\
&G^{(3)}(x_{3},b)=2\epsilon(1+b^2)^{-1-\epsilon}(1+x_{3}^2)^{\epsilon}\biggl(-2\epsilon\int_{+\infty}^{x_{3}}\frac{(1+z^2)^{\epsilon-1}dz}{(1+x_{3}^2)^{\epsilon}}+b-\frac{1+b^2}{b+x_{3}}\biggr)\,.
\end{split}
\end{equation}
It is easy to check that~\eqref{eq:Hdif} is a total differential
\begin{equation}
\frac{\partial F^{(i)}(x_i,b)}{\partial b}=\frac{\partial G^{(i)}(x_i,b)}{\partial x_i}
\end{equation}
as it should be. Then from differential equation $ \frac{\partial \widetilde{H}^{(i)}}{\partial x_i} = F^{(i)},\,i=1,2,3$ we have
\begin{equation}
(-S_{4h})^{\epsilon}\widetilde{H}^{(i)}=\int_{\sign(x_i)\infty}^{x_i}F^{(i)}(z,b)dz+g(b,\epsilon)\,.
\end{equation}
It is easy to check that $ g(b,\epsilon) $ depends only on $ \epsilon $. Indeed,
\begin{equation}
\begin{split}
&\frac{\partial}{\partial b}\Bigl((-S_{4h})^{\epsilon}\widetilde{H}^{(i)}\Bigr)=\int_{\sign(x_i)\infty}^{x_i}\frac{\partial F^{(i)}(z,b)}{\partial b}dz+\frac{\partial g(b,\epsilon)}{\partial b}\\
&\int_{\sign(x_i)\infty}^{x_i}\frac{\partial F^{(i)}(z,b)}{\partial b}dz=\int_{\sign(x_i)\infty}^{x_i}\frac{\partial G^{(i)}(z,b)}{\partial z}dz=G^{(i)}(x_i,b)-G^{(i)}(x_i\rightarrow\sign(x_i)\infty,b)=G^{(i)}(x_i,b)\,,
\end{split}
\end{equation}
where we use asymptotic $ G^{(i)}(x_i\rightarrow\sign(x_i)\infty,b)=2\epsilon b(1+b^2)^{-1-\epsilon} |x_i|^{2\epsilon}\rightarrow0$.  Therefore $ g(b,\epsilon)=g(\epsilon) $. Substituting the explicit form of $ F^{(i)} $ we have 
\begin{equation}\label{eq:HBc}
\begin{split}
& \widetilde{H}^{(1)}=2\epsilon(-S_3)^{-\epsilon}\Re\int_{+\infty}^{x_1}\frac{(1+z^2)^{\epsilon}dz}{b+z+i0}\,,\;\widetilde{H}^{(2)}=2\epsilon(-S_3)^{-\epsilon}\Re\int_{-\infty}^{x_2}\frac{(1+z^2)^{\epsilon}dz}{b+z+i0}\\
&\widetilde{H}^{(3)}=2\epsilon(-S_3)^{-\epsilon}\Re\int_{+\infty}^{x_3}\frac{(1+z^2)^{\epsilon}dz}{b-z+i0}\,.
\end{split}
\end{equation}
Let us integrate the differential equation for $ \widetilde{H}^{(4)} $~\eqref{eq:Hsimpl}. We have
\begin{equation}\label{eq:HBs}
\begin{split}
&\widetilde{H}^{(4)}=-2\epsilon(-S_{4h})^{-\epsilon}\Re\int_{+\infty}^{a}\frac{z^{\epsilon}dz}{z-1+i0}=-2(-t)^{-\epsilon}\biggl(1-\frac{\epsilon}{a}\Re\int_1^{\infty}\frac{z^{\epsilon-1}dz}{z-1/a+i0}\biggr)=\\
&=-2(-t)^{-\epsilon}\Re\,_2F_1\Bigl(1,-\epsilon;1-\epsilon;\frac{1}{a}+i0\Bigr)\,.
\end{split}
\end{equation}
It is easy to see that $ \widetilde{H}_4 $  is expressed by the same function as in the easy box or the box with one off-shell leg. Using equations~\eqref{eq:HBbasis},~\eqref{eq:HBc},~\eqref{eq:HBs} we can write the solution for the hard box integral in the form
\begin{equation}\label{eq:HBf1}
H=\frac{C(\epsilon)}{\epsilon^2s\,t}\biggl(\sum_{i=1}^4\widetilde{H}^{(i)}+g(\epsilon)(-S_{4h})^{-\epsilon}\biggr)\,.
\end{equation}
Let us calculate the constant $ g(\epsilon) $ in the limit $ \Delta_{4h}\rightarrow0 $. We assume that $ (x_1>0,\, x_2<0,\,x_3>0) $
\begin{equation}
t=-\sqrt{\mu_1\mu_2} \,,\;s=-(\sqrt{-\mu_1}-\sqrt{-\mu_2})^2(1+\delta) \,,\;\delta\rightarrow0 \,.
\end{equation}
In the limit $ \delta\rightarrow0 $ we have
\begin{equation}
\widetilde{H}^{(1)}\approx(-\mu_1)^{-\epsilon}\,,\;\widetilde{H}^{(2)}\approx(-\mu_2)^{-\epsilon}\,,\;\widetilde{H}^{(3)}\approx(-s)^{-\epsilon}\,,\;\widetilde{H}^{(4)}\approx-2(-t)^{-\epsilon}\,.
\end{equation}
Therefore from~\eqref{eq:HBf1} we obtain
\begin{equation}
H\approx\frac{C(\epsilon)}{\epsilon^2s\,t}\biggl((-\mu_1)^{-\epsilon}+(-\mu_2)^{-\epsilon}+(-s)^{-\epsilon}-2(-t)^{-\epsilon}+g(\epsilon)\Bigl(\frac{\sqrt{\mu_1\mu_2}}{\delta	}\Bigr)^{-\epsilon}\biggr)\,.
\end{equation}
Since $ \Delta_{4h}=0 $ is not a branch point, then $ g(\epsilon)=0 $\,. Our solution for the hard box integral in the Euclidean region and for $ x_1>0,\,x_2<0,\,x_3>0,\,\Delta_3>0 $ has the following form
\begin{equation}
H=\frac{C(\epsilon)}{\epsilon^2s\,t}\sum_{i=1}^4\widetilde{H}^{(i)}\,.
\end{equation}

\clearpage


\end{document}